\documentclass[preprintnumbers,amsmath,amssymb,aps,prd,longbibliography]{revtex4-1}
\usepackage{amsmath, amssymb}

\usepackage{graphicx}
\usepackage[version=3]{mhchem} 
\usepackage{algorithm}
\usepackage{algorithmic}

\newcommand{\commentoutA}[1]{}

\begin{document}

\preprint{LA-UR-20-22452}

\author{Anders M. N. Niklasson}
\email{amn@lanl.gov}
\affiliation{Theoretical Division, Los Alamos National Laboratory, Los Alamos, New Mexico 87545}

\date{\today}

\title{Density-Matrix based Extended Lagrangian Born-Oppenheimer Molecular Dynamics}

\begin{abstract}
Extended Lagrangian Born-Oppenheimer molecular dynamics
[{\em Phys.\ Rev.\ Lett.\ } {\bf 2008}, {\em 100}, 123004] is presented for Hartree-Fock theory,
where the extended electronic degrees of freedom are represented by a density matrix,
including fractional occupation numbers at elevated electronic temperatures.
In contrast to regular direct Born-Oppenheimer molecular dynamics simulations,
no iterative self-consistent field optimization is required prior to the force evaluations.
To sample regions of the potential energy landscape where the gap is small or vanishing, which leads
to particular convergence problems in regular direct Born-Oppenheimer molecular dynamics simulations,
an adaptive integration scheme for the extended electronic degrees of freedom is presented.
The integration scheme is based on a tunable, low-rank approximation of a fourth-order kernel, ${\cal K}$,
that determines the metric tensor, ${\cal T}\equiv {\cal K}^T{\cal K}$, used in the extended harmonic oscillator of 
the Lagrangian that generates the dynamics of the electronic degrees of freedom.
The formulation and algorithms provide a general guide to implement extended
Lagrangian Born-Oppenheimer molecular dynamics for quantum chemistry,
density functional theory, and semiempirical methods using a density matrix formalism.
\end{abstract}

\maketitle

\section{Introduction}

Extended Lagrangian Born-Oppenheimer molecular dynamics (XL-BOMD) 
\cite{ANiklasson08,PSteneteg10,GZheng11,MCawkwell12,JHutter12,LLin14,MArita14,PSouvatzis14,ANiklasson14,KNomura15,AAlbaugh15,CNegre16,ANiklasson17,JBjorgaard18}
is a general theoretical framework that provides a practical approach to efficient quantum-based molecular dynamics simulations.
XL-BOMD is based on an extended Lagrangian formulation of first principles molecular dynamics
in the spirit of Car-Parrinello molecular dynamics \cite{RCar85,DRemler90,GPastore91,FBornemann98,DMarx00,JHutter12},
where the electronic degrees of freedom are included as extended classical dynamical field variables that are propagated along
the molecular coordinates. In this way the costly overhead of the iterative, self-consistent field (SCF) optimization of the electronic ground state,
which is required in regular direct Born-Oppenheimer molecular dynamics simulation, can be avoided or reduced. 
Various versions and techniques of XL-BOMD have been used and implemented
in a number of software packages, including applications for density functional theory, semiempirical electronic structure theory, 
polarizable force fields, excited state dynamics, and superfluidity 
\cite{ANiklasson11,GZheng11,MCawkwell12,PSouvatzis13, PSouvatzis14,TeraChem,BAradi15,VVitale17,Tinker-HP,LDMPeters17,MArita14,TOtsuka16,THirakawa17,KNomura15,AAlbaugh17,AAlbaugh17b,AAlbaugh18,JBjorgaard18,PHenning19,MKroonblawd19,MKroonblawd20}. 

Originally, XL-BOMD \cite{ANiklasson06,ANiklasson08,ANiklasson09,LLin14} 
was referred to as ``Time-reversible Born-Oppenheimer Molecular Dynamics'', which
often had to be combined with a few self-consistent-field optimization iterations
in each time step to provide stable molecular trajectories. Recent generalized formulations of XL-BOMD
avoid the nonlinearties in the electronic energy expression and the need for any iterative SCF optimization \cite{ANiklasson14,ANiklasson17}. 
It is only these most recent SCF-free formulations of XL-BOMD \cite{ANiklasson14,ANiklasson17,ANiklasson20} that are considered in this article.

To sample regions of the potential energy landscape where the HOMO-LUMO gap is small or vanishing,
which frequently occurs in reactive chemical systems, leads
to particular SCF convergence problems in regular direct Born-Oppenheimer simulations. These regions can also bee a challenge
for XL-BOMD. The main focus of this article is the formulation and implementation of an SCF-free XL-BOMD simulation framework that is using
an adaptive integration scheme for the electronic degrees of freedom, which can be applied also to these more challenging situations.
The electronic structure is described by the density matrix in combination with Hartree-Fock theory and fractional occupation numbers, 
which easily can be adapted to most quantum chemistry codes.

The more recent formulations of XL-BOMD include a metric tensor, ${\cal T} \equiv {\cal K}^T {\cal K}$, in the extended harmonic oscillator that
governs the dynamics of the electronic degrees of freedom.
The tensor is formed by the square of a kernel, ${\cal K}$, that appears in the Euler-Lagrange's equations of motion for the extended electronic degrees of freedom. 
The kernel is given by the inverse of the Jacobian of a residual function, which can be expensive to calculate explicitly in each time step.
Initially the kernel was approximated by a fixed, scaled delta function, ${\cal K} \approx -c \delta({\bf r-r'})$, with $c\in[0,1]$. This approximation
works well for nonreactive chemical systems. However,
for reactive chemical systems, including charge instabilities, the scaled delta-function approximation is often not sufficiently accurate.
The scaled delta-function approximation can be adjusted and improved by a single rank-1 update \cite{ANiklasson17}. 
Unfortunately, this fixed rank-1 adjustment is sometimes also not adequate.
Instead, a tunable rank-$m$ approximation based on a Krylov subspace approximation was proposed \cite{ANiklasson20}. This adjustable
approximation significantly extends the range of systems that can be studied with XL-BOMD simulations. So far, this 
tunable rank-$m$ kernel approximation has been applied only to formulations of XL-BOMD where the extended electronic degrees of 
freedom is described by a vector or a scalar field, for example, the atomic Mulliken net charges in semiempirical electronic structure theory
or the charge density. In these cases the
kernel, ${\cal K}$, and the metric tensor, ${\cal T}$, represent second-order tensor mappings between vectors or scalar fields.  
This is not suitable for many first principles electronic
structure methods. In this article, I will instead present a kernel formalism for XL-BOMD using Hartree-Fock theory 
with a density matrix as the extended electronic degrees of freedom.  Density matrices are used frequently in quantum chemistry methods
as well as in other electronic structure techniques and is therefore often a natural choice for the extended electronic degrees of 
freedom in implementations of XL-BOMD.  In this case we have to use a fourth-order metric
tensor representation of ${\cal T}$ in the harmonic extension of XL-BOMD, where the corresponding kernel, ${\cal K}$, 
represents a fourth-order tensor mapping between matrices. 
The density-matrix adapted methodology requires modifications and it has some limitations compared to the previous formulations. 
The main goal of this article, apart from introducing a higher-order metric tensor formalism for the kernel approximation in the integration
of the extended electornic degrees of freedom, is to provide a general guide how to implement XL-BOMD in electronic software packages using 
an effective single-particle density matrix formalism. This will allow XL-BOMD simulations of a broader class of materials,
including reactive chemical systems.
A computational prototype scheme written in pseudocode, Algorithm 3, summerizes the 
theory and the algorithms.  The discussion is mainly focused on Hartree-Fock theory, but the theory is general and should
be directly applicable also to density functional theory or semiempirical methods using a density matrix formalism.

I will first briefly review regular direct Born-Oppenheimer molecular dynamics using a Hartree-Fock approximation that has been
generalized to finite electronic temperatures using fractional occupation numbers.
Thereafter I present the ideas behind XL-BOMD, its Lagrangian formulation and the equations of motion. 
The integration of the equations of motion is then discussed, in particular, how the fourth-order kernel is approximated in the
integration of the electronic equation of motion. I then present a more complete computational scheme for XL-BOMD in the 
density matrix based Hartree-Fock formalism. At the end I demonstrate some examples before the final discussion and the summary.
A more general description, background and analysis of XL-BOMD can be found in Refs.\ \cite{ANiklasson14,ANiklasson17,ANiklasson20}.

\section{XL-BOMD for Hartree-Fock Theory with Fractional Occupation Numbers}

\subsection{Direct Born-Oppenheimer Molecular Dynamics}

In direct Born-Oppenheimer molecular dynamics based on the spin-restricted Hartree-Fock approximation \cite{Roothaan,RMcWeeny60} that has been
generalized to account for fractional occupation numbers and the electronic free energy at finite electronic temperatures \cite{NMermin63},
the potential energy surface, $U({\bf R})$, can be 
defined through a constrained minimization of a nonlinear density-matrix expression of the free energy, where
\begin{equation}\label{BOPES}
{\displaystyle  U({\bf R})  =   \min_{{\bf D}}   \left\{ 2 {\rm Tr}\left[{\bf h}{\bf D}\right] + {\rm Tr}\left[{\bf D} {\bf G}\left({\bf D}\right)\right] -  2T_e {\cal S}({\bf f}) \right\} 
+  V_{nn}({\bf R})  }
\end{equation}
under the constraints that
\begin{equation}\label{BOPES_Cnstr}
\begin{array}{l}
{\displaystyle  {\bf D} = \sum_i f_i {\bf C}_i {\bf C}_i^T}, \\ 
~~\\
{\displaystyle {\bf C}_i^T{\bf S}{\bf C}_j = \delta_{i,j}},\\
~~\\
{\displaystyle \sum_i f_i = N_{\rm occ}}.
\end{array}
\end{equation}
In this thermal Hartree-Fock formulation ${\bf R} = \{R_I\}$ are the nuclear coordinates (subscript denotes $x$, $y$, and $z$ components of each atom); 
${\bf D} \in {\mathbb{R}}^{N \times N}$ is the constrained single-particle density matrix; ${\bf S} \in \mathbb{R}^{N \times N}$ is the basis-set overlap matrix;
${\bf h} \in \mathbb{R}^{N \times N}$ is the single-electron Hamiltonian matrix; 
$N_{\rm occ}$ is the number of occupied states (each doubly occupied); ${\bf f}$, where $f_i \in [0,1]$, are the fractional occupation numbers; 
$\{{\bf C}_i\}$ is some set of vectors;
$V_{nn}$ is the ion-ion repulsive term; ${\bf G}({\bf D}) \in \mathbb{R}^{N \times N}$ is the electron-electron interaction matrix
corresponding to the Coulomb, ${\bf J}({\bf D}) \in \mathbb{R}^{N \times N}$, and exchange matrix, ${\bf K}({\bf D})  \in \mathbb{R}^{N \times N}$, i.e.\
\begin{equation}\label{J_K}
{\displaystyle {\bf G}({\bf D}) = 2{\bf J}({\bf D}) - {\bf K}({\bf D})};
\end{equation}
$T_e$ is the electronic temperature; and ${\cal S}({\bf f})$ is the electronic entropy contribution, 
\begin{equation}
{\displaystyle {\cal S}({\bf f}) = -k_B \sum_i\left( f_i \ln(f_i) + (1-f_i) \ln(1-f_i)\right),}
\end{equation}
where $k_B$ is Boltzmann's constant. 
At $T_e = 0$ the entropy contribution vanishes and the occupation numbers, $f_i$, are either 1 or 0 as in regular Hartree-Fock theory.
In principle, it is only this ground state solution at $T_e = 0$ that corresponds to a Born-Oppenheimer potential energy surface.
However, because the finite temperature solution in the thermal Hartree-Fock theory is a straightforward generalization, 
I will still call the free energy potential at $T_e > 0$ a Born-Oppenheimer potential energy surface.
The matrices, ${\bf h}$, ${\bf G}({\bf D})$, and ${\bf S}$ in Eq.\ \ref{BOPES} have matrix elements that have been calculated using some
atomic-orbital basis set, $\{\phi_i({\bf r})\}_{i = 1}^N$, for example,
where the overlap matrix, ${\bf S}$, has matrix elements $S_{i,j} = \langle \phi_i\vert \phi_j \rangle, ~~ i,j = 1, 2, \ldots, N $.
The ground state density matrix, ${\bf D}_{\rm min}$, which is attained at the constrained minimum in Eq.\ \ref{BOPES} and
defines the Born-Oppenheimer potential energy surface, $U({\bf R})$, can be calculated as
a solution of the generalized nonlinear algebraic eigenvalue equation,
\begin{equation}\label{NonLinEigV}
{\displaystyle {\bf F}({\bf D}){\bf C}_i = {\bf S}{\bf C}_i\epsilon_i,}
\end{equation}
where ${\bf F}({\bf D})$ is the Hartree-Fock Hamiltonian or Fockian,
\begin{equation}
{\displaystyle {\bf F}({\bf D}) = {\bf h} + {\bf G}({\bf D})  }.
\end{equation}
The density matrix, ${\bf D}$, is then given as a weighted outer product of the eigenstate vectors, $\{{\bf C}_i\}_{i = 1}^N$, where
\begin{equation}
{\displaystyle {\bf D} = \sum_i f_i {\bf C}_i {\bf C}_i^T },
\end{equation}
with the Fermi occupation factors (i.e.\ the fractional occupation numbers) given by
\begin{equation}
{\displaystyle f_i = \left( e^{\beta(\epsilon_i - \mu_0)} + 1 \right)^{-1}},
\end{equation}
where the chemical potential, $\mu_0$, is chosen such that
\begin{equation}\label{focc}
{\displaystyle \sum_i f_i = N_{\rm occ}},
\end{equation}
and $\beta = 1/(k_{\rm B} T_e)$ is the inverse temperature.
The solution of ${\bf D}_{\rm min}$ in the constrained, nonlinear, variational formulation in Eq.\ \ref{BOPES}, given through Eqs.\ \ref{NonLinEigV}-\ref{focc}, 
can be found through an iterative self-consistent-field optimization
procedure, where a new approximate density matrix is updated and mixed with previous approximate solutions 
until the matrix commutation between ${\bf F}({\bf D})$ and ${\bf D}$ vanish at convergence, i.e.\ when ${\bf D} = {\bf D}_{\rm min}$. 
This iterative optimization is often computationally expensive and yet, in practice, never exact.

Once the ground state optimization has been performed, the interatomic forces can be calculated \cite{MWeinert92,RWentzcovitch92,ANiklasson08b} using
the Hellmann-Feynman theorem, i.e.\ using the property that $\partial U({\bf R})/\partial {\bf D} = 0$, which is valid for the exact, fully SCF optimized, 
ground state, ${\bf D}_{\rm min}$.  The equations of motion, 
\begin{equation}
{\displaystyle M_I {\ddot R}_I = -\frac{\partial U({\bf R})}{\partial R_I} },
\end{equation}
where $M_I$ are the atomic masses, can then be integrated. This can be done, for example, by using the time-reversible and
symplectic leapfrog velocity Verlet scheme that then generates the molecular trajectories time step by time step.

The constant of motion is given by the Born-Oppenheimer Hamiltonian,
\begin{equation}
{\displaystyle {\cal H}_{\rm BO} = \frac{1}{2} \sum_I M_I {\dot R}_I^2 + U({\bf R})}.
\end{equation}

\subsection{Some Ideas Behind Density-Matrix based XL-BOMD} \label{Ideas}

By extrapolating the ground state density matrix from previous time steps it is possible to significantly reduce the
computational overhead of the iterative ground state optimization required in Eq.\ \ref{BOPES}.
However, because the variational ground state optimization is approximate and never complete (i.e.\ $\partial U({\bf R})/\partial {\bf D} \approx 0$), 
the forces evaluated using the Hellmann-Feynman theorem are never exactly conservative.  This leads to a systematic drift in the total energy, because of
a broken time-reversal symmetry in the fictitious propagation of the underlying electronic degrees of freedom that is generated through
the extrapolation \cite{DRemler90,PPulay04,ANiklasson07}.  The alternative, to restart the ground state optimization in each new time step
from overlapping atomic densities, preserves the time-reversal symmetry and avoids a systemtic drift in the total energy, but the computational
cost is significantly higher.
Extended Lagrangian Born-Oppenheimer molecular dynamics (XL-BOMD)
\cite{ANiklasson08,PSteneteg10,GZheng11,MCawkwell12,JHutter12,LLin14,MArita14,PSouvatzis14,ANiklasson14,KNomura15,AAlbaugh15,ANiklasson17,JBjorgaard18}
is a framework developed to avoid these shortcomings.

The main idea behind XL-BOMD is based on a backward error analysis. Instead of calculating approximate forces through
and expensive iterative ground-state optimization procedure for an underlying ``exact'' Born-Oppenheimer potential
energy surface, we can calculate exact forces in a fast and simple SCF-free way, but for an underlying approximate {\em shadow}
potential energy surface. In this way we can reduce the computational cost and at the same time restore a consistency between the
calculated forces and the underlying shadow potential.

Density-matrix based XL-BOMD is given in terms of an extended Lagrangian formulation of the dynamics using four approximations: 1) 
The nonlinear density matrix energy expression that is minimized in Eq.\ \ref{BOPES}
is linearized around some approximate density matrix, ${\bf P}$, which is assumed to
be close to the exact ground state, ${\bf D}_{\rm min}$. 
The constrained variational optimization of this linearized energy expression provides the ${\bf P}$-dependent 
ground state density matrix, ${\bf D}[{\bf P}]$, and shadow potential energy surface, which
can be determined in a single SCF-free step without the approximate iterative ground state optimization;
2) The approximate density matrix, ${\bf P}$, around which the linearization is performed, is included
as an extended dynamical tensor variable for a fictitious electronic degrees of freedom that evolves through a harmonic oscillator that is centered
around the best available approximation to the exact ground state density matrix, ${\bf D}_{\rm min}$. 
In this way ${\bf P}$ will remain close to the optimized ground state density;
3) The harmonic potential corresponding to the best available approximation of the ground state density matrix
is given by the generalized square-norm of the residual matrix function, ${\bf D}[{\bf P}]-{\bf P}$, formed by the difference between
${\bf P}$ and the optimized solution of the linearized energy expression, ${\bf D}[{\bf P}]$, using a 
metric tensor, ${\cal T} \equiv {\cal K}^T {\cal K}$, where ${\cal K}$ is the inverse Jacobian of the residual matrix function;
4) The Euler-Lagrange equations of motion are then derived in an adiabatic limit,
where the frequency of the harmonic oscillator extension is assumed to be high compared to the highest frequency of the
nuclear degrees of freedom and the fictitious electron mass parameter goes to zero. 
This classical adiabatic limit corresponds to a Born-Oppenheimer-like approximation for the extended classical
electronic degrees of freedom and leads to a pair of coupled equations of motion 
for the nuclear and the electronic degrees of freedom.

It is only the most recent formulations of XL-BOMD that include the generalized metric tensor in the harmonic oscillator extension 
\cite{ANiklasson14,ANiklasson17,ANiklasson20}. 
The tensor acts such that the dynamical density matrix variable, ${\bf P}$, for the fictitious electronic degrees of freedom, 
oscillate around a much closer approximation to the
exact Born-Oppenheimer ground state compared to the variationally optimized solution of the linearized energy functional, ${\bf D}[{\bf P}]$.
The metric tensor and its various low-rank approximations in the integration of the electronic degrees of freedom
have so far not been discussed explicitly for a density matrix formulation. It has only 
been presented in formulations where the extended electronic degrees of freedom have been
the electron density or the net Mulliken charges, which only requires a metric tensor of second order. 
The main purpose of this article is to present density-matrix based XL-BOMD, which requires a fourth-order metric
tensor, ${\cal T}$, that performs a mapping between matrices.  
I will also use a general atomic-orbital representation of the extended electronic 
degrees of freedom, i.e.\ as in Ref.\ \cite{MArita14}, with a dynamical matrix variable ${\bf X}={\bf P}{\bf S}$ instead of the density matrix ${\bf P}$ or 
its orthogonalized form, ${\bf P}^\perp$. Although a transformation between the different representations
can be performed easily, they are not equivalent in the integration.

The density matrix formulation presented for the thermal Hartree-Fock approximation 
is quite general and should be straightforward to apply to a broad class of 
quantum chemistry techniques, including density functional theory and semiempirical methods.
For density functional theory, the only necessary modification is to replace the Coulomb and exchange matrices, ${\bf G}({\bf P}) = 2{\bf J}({\bf P})-{\bf K}({\bf P})$,
with the corresponding Coulomb matrix, $V_{\rm C}$, and exchange-correlation matrix, $V_{\rm xc}$.

\subsection{Extended Lagrangian}

The theory of XL-BOMD is based on an extended Lagrangian formulation and the four approximations presented in the previous section.
The extended Lagrangian for spin-restricted thermal Hartree-Fock theory, which includes fractional occupation numbers
at finite electronic temperatures, is defined by
\begin{equation}\label{XL}
{\displaystyle {\cal L}({\bf R},{\bf {\dot R}},{\bf X},{\dot {\bf X}}) = \frac{1}{2} \sum_I M_I {\dot R}_I^2  -{\cal U}({\bf R}, {\bf X}) 
+ \frac{\mu}{2} {\rm Tr}\left[\vert {\dot {\bf X}}\vert^2\right]
- \frac{\mu \omega^2}{2} {\rm Tr} \left[ \left({\bf D}[{\bf X}]{\bf S} - {\bf X}\right)^T{\cal T}\left({\bf D}[{\bf X}]{\bf S} - {\bf X}\right) \right]  }.
\end{equation}
Here ${\bf R}$ and ${\bf {\dot R}}$ are the atomic positions and their velocities; ${\bf X} \in \mathbb{R}^{N\times N}$ 
and ${\dot {\bf X}} \in \mathbb{R}^{N \times N}$ are the dynamical matrix variables (where ${\bf X}$ corresponds to ${\bf P}{\bf S}$ as discussed in the previous section) 
that represent the extended electronic degrees of freedom; ${\cal U}({\bf R}, {\bf X})$ is the shadow potential for the electronic free energy at some
electronic temperature $T_e$ that approximates the corresponding exact Born-Oppenheimer free energy surface; ${\cal T} \equiv {\cal K}^T{\cal K}$ is the metric tensor;
$\mu$ is a fictitious electronic mass parameter; and $\omega$ is the frequency of the harmonic oscillator extension.  
${\bf D}[{\bf X}]\in \mathbb{R}^{N\times N}$ is the ${\bf X}$-dependent density matrix given by the constrained variational minimization
of a linearized density matrix function for the electronic energy,
\begin{equation}\label{Dmin}
{\displaystyle {\bf D}[{\bf X}] = \arg \min_{{\bf D}} \left\{ 2 {\rm Tr}\left[{\bf h}{\bf D}\right] + {\rm Tr}\left[\left(2{\bf D}-{\bf X}{\bf S}^{-1}\right) {\bf G}\left({\bf X}{\bf S}^{-1}\right)\right]
 - 2T_e {\cal S}({\bf f}) \right\}  }.
\end{equation}
The same density matrix constraints as in Eq.\ \ref{BOPES_Cnstr} are used and the minimization is performed for the lowest
variationally stationary solution.  It is assumed that ${\bf P} \equiv {\bf X}{\bf S}^{-1} \approx {\bf D}[{\bf X}]$ is an approximate ground state density matrix
and ${\bf G}({\bf P})\equiv {\bf G}({\bf X}{\bf S}^{-1})$ corresponds to the regular Coulomb and exchange matrices, i.e.\ ${\bf G}({\bf P}) = 2{\bf J}({\bf P}) - {\bf K}({\bf P})$.
The dynamical matrix variable ${\bf X}$, in the harmonic extension of the Lagrangian in Eq.\ \ref{XL},
oscillates around an approximate electronic ground state solution represented by ${\bf D}[{\bf X}]{\bf S}$.
The shadow potential energy surface, ${\cal U}({\bf R},{\bf X})$, in Eq.\ \ref{XL} is given by
\begin{equation}\label{ShadowPot}
{\displaystyle {\cal U}({\bf R},{\bf X}) =  2 {\rm Tr}\left[{\bf h}{\bf D}[{\bf X}]\right] 
+ {\rm Tr}\left[\left(2{\bf D}[{\bf X}]-{\bf X}{\bf S}^{-1}\right) {\bf G}\left({\bf X}{\bf S}^{-1}\right)\right] 
- T_e {\cal S}({\bf f})  + V_{nn}({\bf R})},
\end{equation}
which is a close approximation to the fully relaxed ground state Born-Oppenheimer potential energy surface 
if the size of the residual matrix function, ${\bf D}[{\bf X}]{\bf S}-{\bf X}$, is small.

The expression for the harmonic oscillator of the extended Lagrangian in Eq.\ \ref{XL} includes a metric tensor,
\begin{equation}\label{Tensor}
{\displaystyle {\cal T} \equiv {\cal K}^T{\cal K} },
\end{equation}
where ${\cal K} \in \mathbb{R}^{N^2 \times N^2}$ is a kernel that acts as a fourth-order tensor, which performs mappings between matrices. 
This kernel, ${\cal K}$, is defined from the inverse of the Jacobian, ${\cal J}$, of the residual matrix function, where
\begin{equation}\label{Jacobian}
{\displaystyle  {\cal J}_{ij,kl}  = \frac{\partial (\{{\bf D}[{\bf X}]{\bf S}\}_{ij}-X_{ij})}{\partial X_{kl}}, }
\end{equation}
and
\begin{equation}\label{InverseJacobian}
{\displaystyle  {\cal K} = {\cal J}^{-1}. }
\end{equation}

We will never deal with any full fourth-order tensor or its inverse explicitly, only how its low-rank approximations
act on a matrix.  Low-rank approximations are necessary to avoid a large computational overhead even 
for fairly small molecular systems.

\subsection{Variational Optimization of the Density Matrix}

The main computational cost of XL-BOMD is the constrained variational optimization of the density matrix ${\bf D}[{\bf X}]$
in Eq.\ \ref{Dmin}. However, the cost is drastically reduced compared to the nonlinear minimization required in 
direct Born-Oppenheimer molecular dynamics in Eq.\ \ref{BOPES}.
If the approximate Fockian in a nonorthogonal atomic-orbital representation is given by
\begin{equation}
{\displaystyle {\bf F} \equiv {\bf F}({\bf X}) = {\bf h} + {\bf G}({\bf X}{\bf S}^{-1})  },
\end{equation}
with the orthogonalized matrix representation,
\begin{equation}
{\displaystyle {\bf F}^\perp_0 = {\bf Z}^T{\bf F}{\bf Z} , }
\end{equation}
then the constrained density matrix minimization in Eq.\ \ref{Dmin} can be performed by first calculating
\begin{equation}\label{Fermi_DM}
{\displaystyle {\bf D}^\perp[{\bf X}] = \left( e^{\beta({\bf F}^\perp_0 - \mu_0 {\bf I})} + {\bf I} \right)^{-1}},
\end{equation}
where $\mu_0$ is the chemical potential set such that ${\rm Tr}[{\bf D}^\perp[{\bf X}]] = N_{\rm occ}$, followed
by the transform back to the nonorthogonal atomic orbital representation, where
\begin{equation}
{\displaystyle {\bf D}[{\bf X}] = {\bf Z}{\bf D}^\perp[{\bf X}] {\bf Z}^T}.
\end{equation}
Here ${\bf Z} \in \mathbb{R}^{N \times N} $ is the inverse overlap factorization matrix, which is determined by the relation
\begin{equation}\label{Z}
{\displaystyle {\bf Z}^T {\bf S}{\bf Z} = {\bf I}},
\end{equation}
where ${\bf I} \in \mathbb{R}^{N \times N}$ is the identity matrix.
The density-matrix calculation in Eq.\ \ref{Fermi_DM} can be performed using a straightforward matrix diagonalization or by using
a serial Chebyshev or recursive Fermi-operator expansion scheme
\cite{SGoedecker94,RSilver94,RSilver96,AWeisse06,ANiklasson03B,ANiklasson08b,ANiklasson15}.
The diagonalization has a computational complexity that scales cubically, ${\cal O}(N^3)$,
with the number of basis functions, whereas the Fermi-operator expansion methods can be made to
scale only as ${\cal O}(N)$ if we can take advantage of thresholded sparse matrix algebra \cite{SGoedecker99,DBowler12}.
Linear scaling, ${\cal O}(N)$, methods can also be used to calculate ${\bf Z}$ \cite{CNegre16}.

The constrained optimization of ${\bf D}[{\bf X}]$ in Eq.\ \ref{Dmin} is given without any iterative self-consistent
field optimization, because the matrix functional in Eq.\ \ref{Dmin} is linear
in ${\bf D}$. The nonlinear self-consistent field problem in regular Born-Oppenheimer molecular dynamics, which requires
an iterative solution, has been removed. The computational cost of XL-BOMD is therefore significantly
reduced compared to direct Born-Oppenheimer molecular dynamics.
However, ${\bf D}[{\bf X}]$ is not the exact Born-Oppenheimer ground state solution. Nevertheless, it is the exact and
variationally stationary solution of the shadow potential energy surface, ${\cal U}({\bf R}, {\bf X})$, in Eq.\ \ref{ShadowPot},
in the same way as the exact ${\bf D}_{\rm min}$ is the variationally stationary solution for the regular Born-Oppenheimer
potential energy surface, $U({\bf R})$, in Eq.\ \ref{BOPES}.
This simplifies the calculation of interatomic forces that are consistent with
the shadow potential energy surface and we can avoid 
contributions from terms including $\partial {\bf D}[{\bf X}]/\partial R_I$, since
\begin{equation}
{\displaystyle \frac{\partial {\cal U}({\bf R},{\bf X})}{\partial {\bf D}[{\bf X}]} \frac{\partial {\bf D}[{\bf X}]}{\partial R_I} = 0}.
\end{equation}
With XL-BOMD we are thus able to calculate accurate conservative forces that are consistent with the shadow potential.
This is achieved without requiring any costly iterative self-consistent field optimization procedure prior to the force evaluations.
Moreover, as long as ${\bf X}{\bf S}^{-1}$ is a reasonably close approximation to the exact self-consistent ground state density, ${\bf D}_{\rm min}$, i.e.\ as long as
the residual matrix function ${\bf D}[{\bf X}]{\bf S} - {\bf X}$ is small, the shadow potential energy surface, ${\cal U}({\bf R},{\bf X})$, in Eq.\ \ref{ShadowPot},
is close to the exact Born-Oppenheimer potential, $U({\bf R})$, \cite{ANiklasson17}.

\subsection{Equations of Motion}

Atoms are moving on a slow time scale compared to the electronic motion.
If the electrons are in the ground state initially we may therefore assume they will evolve adiabatically
along the electronic ground state as the atoms are moving. This is the reasoning behind the Born-Oppenheimer approximation in
quantum-based molecular dynamics simulations \cite{WHeitler27,MBorn27,DMarx00}.
In the derivation of the equations of motion from the Euler-Lagrange equations for the extended Lagrangian in Eq.\ \ref{XL}
we can also apply an adiabatic approximation that separates the motion between the nuclear and the extended electronic degrees of freedom.
In this classical adiabatic approximation, we assume that $\omega$ that determines
the frequency of the extended electronic degrees of freedom, ${\bf X}(t)$ and ${\dot {\bf X}}(t)$, is large compared
to some highest frequency, $\Omega$, of the nuclear degrees of freedom, ${\bf R}(t)$ and ${\bf {\dot R}}(t)$. 
In the derivation of the Euler-Lagrange equations of motion
of the Lagrangian in Eq.\ \ref{XL} we then let $\omega/\Omega \rightarrow \infty$, such that
$\mu \omega = {\rm constant}$. In this classical adiabatic ``Born-Oppenheimer-like'' limit \cite{ANiklasson14,ANiklasson17},
for which $\mu \rightarrow 0$, it can be shown that the size of the residual ${\bf D}[{\bf X}]{\bf S}-{\bf X}$ scales as $1/\omega^{2}$
and that $\partial {\cal U}({\bf R},{\bf X})/\partial {\bf X}$ scales linearly with the residual ${\bf D}[{\bf X}]{\bf S}-{\bf X}$.  The
equations of motion for XL-BOMD in this adiabatic ``mass-zero'' limit are then given by
\begin{equation}\label{EqR}
{\displaystyle  M_I {\ddot R}_I = - \left. \frac{\partial {\cal U}({\bf R},{\bf X})}{\partial R_I}\right \vert_{\bf X},   }
\end{equation}
for the nuclear degrees of freedom and 
\begin{equation}\label{EqX}
{\displaystyle  {\ddot {\bf X}} = - \omega^2 {\cal K} \left({\bf D}[{\bf X}]{\bf S} - {\bf X}\right), }
\end{equation}
for the electronic degrees of freedom.
The corresponding constant of motion is given by the shadow Hamiltonian,
\begin{equation}\label{ConstH} \begin{array}{l}
{\displaystyle {\cal H}_{\rm XL-BOMD} = \frac{1}{2} \sum_I M_I {\dot R}^2_I + {\cal U}({\bf R},{\bf X}) }\\
~~\\
{\displaystyle  ~~  = \frac{1}{2} \sum_I M_I {\dot R}^2_I + 2 {\rm Tr}\left[{\bf h}{\bf D}[{\bf X}]\right]
 + {\rm Tr}\left[\left(2{\bf D}[{\bf X}]-{\bf X}{\bf S}^{-1}\right) {\bf G}\left({\bf X}{\bf S}^{-1}\right)\right] 
- T_e {\cal S}({\bf f})  + V_{nn}({\bf R})}.
\end{array}
\end{equation}
Equations \ref{EqR} and \ref{EqX} together with the constant of motion in Eq.\ \ref{ConstH} are the three central equations
that governs the dynamics of XL-BOMD for thermal Hartree-Fock theory.

\section{Integrating XL-BOMD}

Integrating the equations of motion, Eqs.\ \ref{EqR} and \ref{EqX}, includes some technical challenges.
One problem is to keep the extended electronic degrees of freedom synchronized with the nuclear degrees of freedom.
In a perfectly time-reversible integration, numerical noise will accumulate, which eventually leads to a divergence 
between the exact Born-Oppenheimer ground state that is determined by the nuclear coordinates and the
electronic degrees of freedom around which the density matrix energy function is linearized. ${\bf P} \equiv {\bf X} {\bf S}^{-1}$ is then
no longer a close approximation to the ground state, ${\bf D}_{\rm min}$, and the accuracy of the linearized functional
in Eq.\ \ref{Dmin} breaks down.  Another problem
is to approximate the inverse Jacobian kernel, ${\cal K}$, in the electronic equations of motion. The full exact kernel
defined in Eqs.\ \ref{Jacobian} and \ref{InverseJacobian} 
is expensive (in practice impossible) to calculate explicitly and its action on the residual ${\bf D}[{\bf X}]{\bf S}-{\bf X}$ has to be approximated in the integration 
of the electronic degrees of freedom, Eq.\ \ref{EqX}.

\subsection{Modified Verlet Scheme with Damping}

For the integration of the combined nuclear and electronic degrees of freedom in Eqs.\ \ref{EqR} and \ref{EqX}
we can use a modified leapfrog velocity Verlet scheme
\cite{ANiklasson09,PSteneteg10,GZheng11} that includes an additional dissipative term in the integration of the extended
electronic degrees of freedom. This additional term breaks the time-reversal symmetry to
some chosen higher odd-order in the integration time step, $\delta t$, which dampens the accumulation of numerical noise
that otherwise could cause instabilities in a perfectly reversible integration. In this
way the evolution of the electronic degrees of freedom stays synchronized to the dynamics of the nuclear motion.
The modified leapfrog velocity Verlet integration scheme for the integration
of the nuclear and electronic degrees of freedom is given by
\begin{equation}\label{Integration}\begin{array}{l}
{\displaystyle {{\dot {\bf R}}}(t + \frac{\delta t}{2})  = {{\dot {\bf R}}}(t) + \frac{\delta t}{2} {{\ddot {\bf R}}}(t)},\\
~~\\
{\displaystyle {{\bf R}}(t + \delta t) = {{\bf R}}(t) + \delta t {{\dot {\bf R}}}(t + \frac{\delta t}{2})},\\
~~\\
{\displaystyle {\bf X}(t + \delta t) = 2{\bf X}(t) - {\bf X}(t - \delta t) + \delta t^2 {\ddot {\bf X}}(t) 
+ \alpha \sum_{k = 0}^{k_{\rm max}} c_k {\bf X}(t-k \delta t)},\\
~~\\
{\displaystyle {{\dot {\bf R}}}(t + \delta t) = {{\dot {\bf R}}}(t + \frac{\delta t}{2})
+ \frac{\delta t}{2} {{\ddot {\bf R}}}(t+\delta t)}.\\
\end{array}
\end{equation}
The last term in the integration of ${\bf X}(t$) is the additional damping term,
where the coefficients, $\alpha$ and $\{ c_k \}_{k = 0}^{k_{\rm max}}$,
as well as a dimensionless constant, $\kappa = \delta t^2 \omega^2$,
have been optimized for various values of $k_{\rm max}$ and are given in Ref.\ \cite{ANiklasson09}.
In the initial time step ${\bf X}(t_0-k\delta t)$ for $k = 0,1,\ldots,k_{\rm max}$ are all set to the
fully converged regular Born-Oppenheimer ground state density, ${\bf D}_{\rm min}$, times the overlap matrix ${\bf S}$, i.e.\ 
at $t_0$ we set ${\bf X}(t_0-k\delta t) = {\bf D}_{\rm min}{\bf S}$ for $k = 0,1,\ldots,k_{\rm max}$.
A reasonably well-converged iterative self-consistent field optimization is thus required, 
but only in the first initial time step.  The modified Verlet integration scheme works similar to a Langevin dynamics, 
but where the stochastic term is generated by the intrinsic numerical noise
of the system instead of an external random number generator
and where the dissipation is given by a higher-order time-derivative term instead of a first-order
velocity-driven friction term.
In general, this solution works well without any significant drift in the constant of motion on time scales
relevant for quantum-based Born-Oppenheimer molecular dynamics.
However, a number of alternative integration schemes for XL-BOMD have
been proposed and analyzed \cite{AOdell09,AOdell11,AAlbaugh15,VVitale17,AAlbaugh17,AAlbaugh18}.

\subsection{Krylov Subspace Approximation of the Inverse Jacobian Kernel}

The integration of the  electronic degrees of freedom
in Eq.\ \ref{Integration} requires the calculation of the second-order time derivative of ${\bf X}(t)$, which is
given by the electronic equations of motion in Eq.\ \ref{EqX}.
This equation includes the inverse Jacobian kernel, ${\cal K} \in \mathbb{R}^{N^2 \times N^2}$, 
that is given from the Jacobian of the residual matrix function $({\bf D}[{\bf X}]{\bf S} - {\bf X}) : {\mathbb R}^{N \times N} \rightarrow {\mathbb R}^{N \times N}$ using
Eqs.\ \ref{Jacobian} and \ref{InverseJacobian}. A technique to perform a tunable and adaptive 
approximation of the kernel for the integration
of the electronic degrees of freedom in XL-BOMD was developed \cite{ANiklasson20},
but for a residual function ${\bf f}({\bf n}) = ({\bf q}[{\bf n}]-{\bf n}): \mathbb{R}^{N} \rightarrow \mathbb{R}^N$. That method was derived from an expression
of the Jacobian of ${\bf f(n)}$, which is based on a set of general directional derivatives, 
\begin{equation}
{\displaystyle {\bf f}_{{\bf v}_i} = \left. \frac{\partial {\bf f}({\bf n} + \lambda {\bf v}_i)}{\partial \lambda}\right \vert_{\lambda = 0}},
\end{equation}
along directions ${\bf v}_i \in {\mathbb R}^N$ instead of partial derivatives, $\partial {\bf f(n)}/\partial n_i$, 
with respect to the individual components of ${\bf n}$. The Jacobian, ${\bf J} \in \mathbb{R}^{N \times N}$, 
in this case, can then be approximated by a rank-$m$ expression, 
\begin{equation}
{\displaystyle  {\bf J} \approx \sum_{i,j = 1}^m {\bf f}_{{\bf v}_i} L_{ij} {\bf v}_j^T },
\end{equation}
where ${\bf L} = {\bf O}^{-1}$, which is the inverse overlap with matrix elements, $O_{ij} = {\bf v}_i^T{\bf v}_j$.
The directional derivatives are calculated with quantum perturbation theory and the directions $\{{\bf v}_i\}_{i=1}^m$ are chosen
from a Krylov subspace approximation \cite{ANiklasson20}. The kernel is then determined by a psedoinverse 
of the low-rank Jacobian. In combination with preconditioning, this provides a rapidly converging
low-rank approximation of the kernel in the
integration of the  electronic degrees of freedom. The method is directly related to a single step in 
so-called Jacobian-free Newton Krylov methods used in the iterative solution of systems of nonlinear 
equations \cite{DAKnoll04} and the corresponding residual minimization methods \cite{DGAnderson65,YSaad86,YSaad96}.

Here we need to adapt the previos approximation method of the Jacobian and its pseudoiverse \cite{ANiklasson20} to the
residual matrix function, $({\bf D}[{\bf X}]{\bf S}-{\bf X}) : {\mathbb R}^{N \times N} \rightarrow {\mathbb R}^{N \times N}$.
We start by rewriting Eq.\ \ref{EqX} in a more general but equivalent
form, using a preconditioner, ${\cal K}_0$, i.e.\ where
\begin{equation}\label{Xddot}
{\displaystyle  {\ddot {\bf X}} = - \omega^2 ({\cal K}_0{\cal J})^{-1}{\cal K}_0 \left({\bf D}[{\bf X}]{\bf S} - {\bf X}\right). }
\end{equation}
In the ideal case we assume that ${\cal K}_0 {\cal J} \approx {\cal I}$, where ${\cal I}$ is the fourth-order identity tensor. 
We will then try to
find a low-rank approximation of how the preconditioned kernel, $({\cal K}_0{\cal J})^{-1}$,
acts on the modified residual, ${\cal K}_0 \left({\bf D}[{\bf X}]{\bf S} - {\bf X}\right)$. For the second-order kernels it is possible to
construct efficient preconditioners by a direct full calculation of the kernel at the initial time step,
which then can be kept as a constant preconditioner during the molecular dynamics simulation. 
This preconditioning technique generates rapidly converging low-rank approximations of the kernel.
With the fourth-order kernel this direct method is no longer possible, except for very small molecular systems.  I will therefore not 
pursue the use of preconditioners in this article, even if they can be constructed in different forms \cite{NiklassonUnPub}. 
I will still keep a ${\cal K}_0$ in the discussion, but only to keep it general.
It will be demonstrated in the examples, that we can achieve accurate low-rank approximations even without preconditioner, 
i.e.\ when the identity is used as a preconditioner and ${\cal K}_0 = {\cal I}$. Another difference to the original Krylov subspace approximation is that
the metric of the inner product has to be modified. Instead of a vector-product, $\langle {\bf v}_i,{\bf v}_j \rangle = {\bf v}_i^T {\bf v}_j$,
where ${\bf v}_i \in \mathbb{R}^N$, we will use the matrix generalization, $\langle {\bf V}_i,{\bf V}_j \rangle = {\rm Tr}[{\bf V}_i^T {\bf V}_j]$,
where ${\bf V}_i \in \mathbb{R}^{N \times N}$.  The matrix norm, $\|{\bf V}_i\| = \sqrt{\langle {\bf V}_i,{\bf V}_i \rangle} = \sqrt{{\rm Tr}[{\bf V}_i^T {\bf V}_i]}$,
then corresponds to the Frobenius norm. The general principle for the original formulation can then be kept.
With this matrix generalization of the method in Ref.\ \cite{ANiklasson20}, the rank-$m$ Krylov 
subspace approximation for how the preconditioned kernel acts on the modified matrix residual function is given by
\begin{equation} \label{KernelAppr}
{\displaystyle ({\cal K}_0{\cal J})^{-1}{\cal K}_0 \left({\bf D}[{\bf X}]{\bf S} - {\bf X}\right) \approx \sum_{i,j = 1}^m {\bf V}_i M_{ij} \langle {\bf W}_j,{\bf W}_0\rangle  }.
\end{equation}
The algorithm that generates the matrices ${\bf V}_i \in \mathbb{R}^{N \times N}$, ${\bf M} \in \mathbb{R}^{m \times m}$, 
and ${\bf W}_i \in \mathbb{R}^{N \times N}$, is given in Algorithm 1. 
It is an adaptive scheme, where the order $m$ of the rank-$m$ approximation is tunable by a chosen tolerance level.
With this tunable rank-$m$ Krylov subspace approximation we can then integrate the electronic degrees of freedom as
\begin{equation}\label{IntEqX}
{\displaystyle {\bf X}(t+\delta t) = 2{\bf X}(t) - {\bf X}(t-\delta t) 
 - \delta t^2 \omega^2 \sum_{i,j=1}^m {\bf V}_i M_{i,j} \langle {\bf W}_j, {\bf W}_0\rangle 
 + \alpha \sum_{k=0}^{k_{\rm max}} c_k {\bf X}(t-k\delta t)}.
\end{equation}

Apart from the low-rank approximation of the inverse Jacobian kernel acting on the residual function in Eq.\ \ref{KernelAppr}, with the matrices constructed as in
Algorithm 1, there are a number of alternative low-rank approximations that can be used. These techniques are discussed in the appendix of Ref.\ \cite{ANiklasson20}.
Those methods could also be adjusted to the density matrix formalism in analogy to the technique presented here. However, here
I will only consider the rank-$m$ approximation in Eq.\ \ref{KernelAppr} and Algorithm 1.

\begin{algorithm}[H]
\label{KernelApproximation}
\caption{This algorithm generates $\{{\bf V}_i\}$, ${\bf M}$, and $\{{\bf W}_i\}$ for the kernel approximation, Eq.\ \ref{KernelAppr}, 
in the integration of the electronic degrees of freedom, Eqs.\ \ref{EqX} and \ref{IntEqX}, using
a preconditioned rank-$m$ Krylov subspace approximation of the inverse Jacobian kernel, where
$({\cal K}_0 {\cal J})^{-1}{\cal K}_0({\bf D}[{\bf X}]{\bf S} - {\bf X}) \approx \sum_{i,j}^m {\bf V}_i M_{ij} \langle {\bf W}_j, {\bf W}_0\rangle$. 
The algorithm is based on the Krylov subspace approximation of the inverse Jacobian kernel presented in Ref.\ \cite{ANiklasson20}.
The inner product is given by $\langle {\bf V}_i, {\bf V}_j \rangle = {\rm Tr}[{\bf V}_i^T {\bf V}_j]$, with the Frobenius matrix norm $\| {\bf V}_i\| = \sqrt{\langle {\bf V}_i, {\bf V}_i \rangle}$.
The trace conserving canonical density matrix perturbation, $\left. \partial_\lambda {\bf D}[{\bf F}_0({\bf X})+\lambda {\bf F}_1({\bf V}_m)] \right \vert _{\lambda = 0}$, can be
performed with Algorithm 2.}
\algsetup{indent=1em}
\begin{algorithmic}
\STATE $ {\bf W}_0 = {\cal K}_0({\bf D}[{\bf X}] - {\bf X}), ~~ m = 0$
\WHILE{Error  $ >$ Chosen Tolerance}
 \STATE $ m = m + 1$
 \STATE $ {\bf V}_m = {\bf W}_{m-1}$
 \STATE $ {\bf V}_m = {\bf V}_m - \sum_{j = 1}^{m-1}\langle {\bf V}_i, {\bf V}_j \rangle {\bf V}_j$
 \STATE $ {\bf V}_m = {\bf V}_m \|{\bf V}_m\|^{-1}$
 \STATE $ {\bf W}_{m} =  {\cal K}_0\left( \left. \partial_\lambda {\bf D}[{\bf F}_0({\bf X})+\lambda {\bf F}_1({\bf V}_m)] \right \vert _{\lambda = 0} - {\bf V}_m\right)$
 \STATE $ O_{ij} = \langle {\bf W}_i, {\bf W}_j \rangle, ~i,j = 1,2,\ldots, m$
 \STATE $ {\bf M} = {\bf O}^{-1} $
 \STATE Error = $\left \|\left(\sum_{i,j = 1}^m {\bf W}_i M_{ij} \langle {\bf W}_j, {\bf W}_0\rangle - {\bf W}_0\right)\right \|/\|{\bf W}_0\| $
\ENDWHILE
 \STATE $\Rightarrow ({\cal K}_0{\cal J})^{-1}{\cal K}_0 \left({\bf D}[{\bf X}]{\bf S} - {\bf X}\right) \approx \sum_{i,j}^m {\bf V}_i M_{ij} \langle {\bf W}_j,{\bf W}_0\rangle$
\end{algorithmic}
\end{algorithm}

The calculation of ${\bf W}_m$ in Algorithm 1 requires the directional derivatives of the density matrix,
\begin{equation}
{\displaystyle   \left. {\partial}_\lambda {\bf D}[{\bf F}_0({\bf X})+\lambda {\bf F}_1({\bf V}_m)] \right \vert _{\lambda = 0}
 = \left. {\bf Z} \left( \frac{\partial}{\partial \lambda} {\bf D}^\perp [{\bf F}^{\perp}_0+\lambda {\bf F}^{\perp}_1({\bf V}_m)]\right \vert _{\lambda = 0} \right){\bf Z}^T  },
\end{equation}
where
\begin{equation}
{\displaystyle  {\bf F}_0^\perp = {\bf Z}^T {\bf F}({\bf X}) {\bf Z}  }
\end{equation}
with
\begin{equation}
{\displaystyle  {\bf F}({\bf X}) = {\bf h} + {\bf G}({\bf X}{\bf S}^{-1})  }
\end{equation}
and
\begin{equation}
{\displaystyle  {\bf F}_1^\perp({\bf V}_m) = {\bf Z}^T {\bf G}({\bf V}_m {\bf S}^{-1}) {\bf Z}  }.
\end{equation}
The directional density matrix derivative,
\begin{equation}
{\displaystyle {\bf D}_1^\perp = \left. \frac{\partial}{\partial \lambda} {\bf D}^\perp [{\bf F}^{\perp}_0+\lambda {\bf F}^{\perp}_1({\bf V}_m)]\right \vert _{\lambda = 0}},
\end{equation} 
corresponds to a calculation of the first-order (traceless) response,
\begin{equation}\label{DMResponse}
{\displaystyle   {\bf D}_1^\perp = \left. \frac{\partial}{\partial \lambda} \left( e^{\left[\beta\left({\bf F}_0^\perp - \mu_0 {\bf I} + \lambda ({\bf F}_1^\perp - \mu_1 {\bf I}) \right) \right]} + {\bf I}\right)^{-1} \right \vert_{\lambda = 0}},
\end{equation}
with respect to the perturbation ${\bf F}_1^\perp$ in the Fockian.
The chemical potential $\mu_0$ in the density matrix derivative in Eq.\ \ref{DMResponse} is assumed to be known
and the first-order response in the chemical potential, $\mu_1$, is chosen such that
\begin{equation}
{\displaystyle  {\rm Tr}\left[{\bf D}_1^\perp\right] = 0}.
\end{equation}
This traceless density matrix response can be calculated using canonical density matrix perturbation theory \cite{ANiklasson15}.
An algorithm to calculate the density matrix response ${\bf D}_1^\perp$, which is based 
on a simplified version of the canonical density matrix perturbation algorithm in Ref.\ \cite{ANiklasson15},
using a diagonal eigenbasis representation and assuming a known chemical potential $\mu_0$, is given in Algorithm 2. 
It is constructed from a combination of a recursive Fermi-operator expansion scheme 
\cite{ANiklasson03B,ANiklasson08b} for fractional occupation numbers and density matrix perturbation theory \cite{ANiklasson04}.
Originally the chemical potential, $\mu_0$, and the chemical potential derivative, $\mu_1$, were assumed to be unknown and instead
updated iteratively using a Newton-Raphson scheme based on linearizations of the density matrix and the density matrix response
around approximate $\mu_0$ and $\mu_1$. With an exact $\mu_0$ known initially, the linearization of the density matrix
response with respect to $\mu_1$ is exact in a single step and no iterative adjustments are necessary.
An alternative response calculation that is more similar to regular Rayleigh-Schr\"{o}dinger perturbation theory 
for systems with fractional occupation numbers \cite{SLAdler62,NWiser63},
which also can be adjusted in the same way for the response in the chemical potential $\mu_1$ as in Algorithm 2,
was recently proposed by Nishimoto \cite{YNishimoto17}.

\newpage

\begin{algorithm}[H]
\label{FirstOrderDMResponse}
\caption{Simplified version of the canonical density matrix perturbation algorithm in Ref.\ \cite{ANiklasson15}
in a diagonal matrix representation for first-order response calculations, assuming a known chemical potential $\mu_0$ and a molecular orbital
eigenbasis representation with eigenvectors, ${\bf Q}$, of ${\bf F}_0^\perp$.
All matrices with superscript $^{(0)}$ are diagonal.
Dominating ${\cal O}(N^3)$ cost is marked by two asterisks (*), where $N$ is the matrix dimension.
Input: $\mu_0$, ${\bf Q}$, ${\bf E}$, $n$, $\beta$ or $T_e$, ${\bf F}_0^\perp$, and ${\bf F}_1^\perp$ (${\bf F}_0^\perp$ and ${\bf F}_1^\perp$ in orthogonalized basis-set representation); Output: ${\bf D}_1^\perp = \left. \frac{\partial}{\partial \lambda} {\bf D}^\perp [{\bf F}^{\perp}_0+\lambda {\bf F}^{\perp}_1]\right \vert _{\lambda = 0}$ ; Optional Output: ${\bf D}_0 = {\bf Q}{\bf X}^{(0)}{\bf Q}^T$}
\algsetup{indent=1em}
\begin{algorithmic}
\STATE $ \mu_0 ~~ \mbox{Chemical potential of the unperturbed system}$
\STATE $ {\bf Q} ~~ \mbox{Matrix of eigenvectors of the unperturbed system, i.e.\ where } {\bf Q}^T{\bf F}^\perp_0{\bf Q} = {\bf E}$
\STATE $ {\bf E} ~~ \mbox{Diagonal matrix of eigenvalues of the unperturbed system}$
\STATE $ n ~~ \mbox{Number of recursion steps},~ \mbox{typically} ~n \in [6,10]$
\STATE $ \beta = 1/(k_B T_e) ~~ \mbox{Inverse temperature}$
\STATE $ c = 2^{-2-n}\beta ~~\mbox{Initial normalization constant}$
\STATE $ {\bf F}^{(0)} = {\bf Q}^T{\bf F}_0^\perp {\bf Q}  = {\bf E} ~~\mbox{Unperturbed Fockian in the diagonal eigenvalue representation}$
\STATE $ {\bf F}^{(1)} = {\bf Q}^T{\bf F}_1^\perp {\bf Q} ~~\mbox{Perturbed Fockian in molecular orbital representation (*)}$
\STATE $ {\bf X}^{(0)} = \frac{1}{2} {\bf I} - c({\bf F}^{(0)}-\mu_0 {\bf I}) ~~\mbox{Initialization}$
\STATE $ {\bf X}^{(1)} = - c{\bf F}^{(1)} ~~\mbox{Initialization}$
\FOR{i = 1, n}
 \STATE $ {\bf X}^{(1)}_{\rm tmp} = {\bf X}^{(0)} {\bf X}^{(1)} + {\bf X}^{(1)}{\bf X}^{(0)} ~~ \mbox{Diagonal}\times \mbox{full matrix multiplications}$
 \STATE $ {\bf Y}^{(0)}_{\rm tmp} = \left( 2{\bf X}^{(0)}\left({\bf X}^{(0)} - {\bf I}\right)+{\bf I}\right)^{-1} ~~ \mbox{Diagonal matrix inversion}$
 \STATE $ {\bf X}^{(0)} = {\bf Y}^{(0)}_{\rm tmp} {\bf X}^{(0)}{\bf X}^{(0)} ~~\mbox{Diagonal}\times \mbox{diagonal matrix multiplications}$
 \STATE $ {\bf X}^{(1)} = {\bf Y}^{(0)}_{\rm tmp}\left({\bf X}^{(1)}_{\rm tmp} + 2({\bf X}^{(1)}-{\bf X}^{(1)}_{\rm tmp}){\bf X}^{(0)}\right) ~~\mbox{Diagonal}\times
\mbox{full matrix multiplications}$
\ENDFOR
\STATE $ {\bf D}^{(0)}_\mu = \beta {\bf X}^{(0)}({\bf I}-{\bf X}^{(0)})  ~~\mbox{Diagonal density matrix derivative } d{\bf D}^{(0)}/d\mu$
\STATE $ \mu_1 =  - {\rm Tr}[{\bf X}^{(1)}]/{\rm Tr}[{\bf D}^{(0)}_\mu] ~~\mbox{First-order response in } \mbox{ from } \lambda {\bf F}^{(1)}$
\STATE $ {\bf D}^{(1)} = {\bf X}^{(1)} +  {\bf D}^{(0)}_\mu \mu_1  ~ \mbox{Canonical (traceless) first-order density matrix response}$
\STATE $ {\bf D}_1^\perp = {\bf Q}{\bf D}^{(1)}{\bf Q}^T ~~\mbox{First-order response back to the original orthogonal basis-set representation (*)}$
\end{algorithmic}
\end{algorithm}

Algorithm 1 and 2 represent a practical approach to integrate the electronic equations of motion in XL-BOMD as in Eq.\ \ref{IntEqX}
using a density matrix formalism for the extended electronic degrees of freedom. If the diagonalization of ${\bf F}^\perp_0$ has
been performed, no additional diagonalization is needed in the rank-$m$ Krylov subspace approximation of the kernel.
This means that the computational overhead of each extra rank in the low-rank approximation has a limited overhead
compared to a full self-consistent field iteration, which also includes a diagonalization of the Fockian.
A summary in terms of a computational prototype scheme for density-matrix based XL-BOMD is
expressed in a pseudocode given in the section below.

\section{Computational Scheme for Density-Matrix Based XL-BOMD} \label{XL-BOMD_Scheme}

To provide a practical guide for the implementations that is suitable for density-matrix based electronic structure codes,
I present a detailed computational scheme for the XL-BOMD framework using the thermal Hartree-Fock theory 
using a density matrix formalism. The scheme can be generalized and applied also to other electronic structure methods where the 
electronic degrees of freedom can be represented by the single-particle density matrix with fractional occupation numbers.
The XL-BOMD scheme in presented with a pseudocode in Algorithm 3 and represents a comprehensive summary of this article. 

The algorithm starts after an initial full self-consistent field optimization of the
ground state density matrix, ${\bf D}_{\rm min}$, has been performed, which is attained at the constrained variational minimum in Eq.\ \ref{BOPES}.
Here ${\bf G}({\bf R},{\bf P}) = 2{\bf J}({\bf P}) - {\bf K}({\bf P})$ are the Coulomb and exchange matrices in restricted Hartree-Fock theory calculated at the atomic configuration, ${\bf R}$,
where ${\bf P} \equiv {\bf X}{\bf S}^{-1}$.  ${\cal U}({\bf R},{\bf X})$ is the
shadow potential energy surface including the electronic entropy at an electronic temperature $T_e$, as given in Eq.\ \ref{ShadowPot}.
The electronic temperature is given by the inverse temperature $\beta = 1/(k_{\rm B}T_e)$, where $k_{\rm B}$ is the Boltzmann constant.

The XL-BOMD simulation starts after an initialization, where ${\bf X}(t_0-k\delta t) = {\bf D}_{\rm min}{\bf S}, ~~ k = 0,1,2,\ldots,k_{\rm max}$ for
some given configuration ${\bf R}_0$ with velocities ${\bf{\dot  R}}_0$.
The modified Leapfrog velocity Verlet integration scheme, Eq.\ \ref{Integration}, is used
with the coefficients $\{c_k\}$, $\alpha$ and $\kappa = \delta t^2 \omega^2$ given
in Ref.\ \cite{ANiklasson09}. The degrees of freedom are the atomic positions ${\bf R}(t)$,
their velocities ${\bf \dot {R}}(t)$, and the electronic degrees of freedom, ${\bf X}(t)$, which approximately corresponds
to the product of an approximate density matrix and the overlap matrix.
The matrix inner product, $\langle {\bf X},{\bf Y} \rangle = {\rm Tr}[{\bf X}^T {\bf Y}]$, with the corresponding
Frobenius matrix norm are used, where $\|{\bf X}\| = \sqrt{{\rm Tr}[{\bf X}^T{\bf X}]}$. The orthogonalized rank-$m$ Krylov subspace approximation of the
kernel ${\cal K}$ in the integration of the electronic degrees of freedom, Eq.\ \ref{Xddot}-\ref{IntEqX}, based on Ref.\ \cite{ANiklasson20} is used.
The force evaluations are based on
\begin{equation}
{\displaystyle \frac{\partial {\cal U}({\bf R},{\bf X})}{\partial R_I} = 2{\rm Tr}[{\bf h}_{R_I} {\bf D}] + {\rm Tr}[(2{\bf D}-{\bf P}){\bf G}_{R_I}] + \partial V_{nn}/\partial R_I - 2 {\rm Tr} [{\bf Z}{\bf Z}^T{\bf F}{\bf D} {\bf S}_{R_I}]},
\end{equation}
and include terms like 
\begin{equation}
{\displaystyle {\bf h}_{R_I} = \frac{\partial {\bf h}}{\partial R_I}},
\end{equation}
\begin{equation}
{\displaystyle {\bf S}_{R_I} = \frac{\partial {\bf S}}{\partial R_I}},
\end{equation}
\begin{equation}
{\displaystyle {\bf G}_{R_I} = \left.  \frac{\partial {\bf G}({\bf R},{\bf P})}{\partial R_I}\right\vert_{{\bf P}} },
\end{equation}
and the Pulay force-term \cite{PPulay69},
\begin{equation}
{\displaystyle f_{\rm Pulay} = - 2 {\rm Tr} [{\bf Z}{\bf Z}^T{\bf F}{\bf D} {\bf S}_{R_I}]},
\end{equation}
which has been adapted for fractional occupation numbers \cite{MWeinert92,RWentzcovitch92,ANiklasson08b}.
The inverse overlap factorization matrix, ${\bf Z}$, is chosen either as the inverse square root of the overlap matrix, i.e.\ ${\bf Z} = {\bf S}^{-1/2}$,
or some other matrix for which ${\bf Z}^T{\bf S}{\bf Z}={\bf I}$ \cite{ANiklasson04B}. The quantum response calculation of ${\bf D}_1^\perp$ can be performed with Algorithm 2.

Apart from the construction of the the Coulomb and exchange matrix, ${\bf G}({\bf R},\delta {\bf D})$, the computational cost for each new rank update 
in Algorithm 3 is dominated by the response calculation and its required transformations into the molecular eigenbasis and back to the atomic-orbital representation.
If the molecular eigenbasis ${\bf Q}$ and inverse factorization matrix ${\bf Z}$ in Algorithm 2 and Algorithm 3 are combined, the cost of
each response calculation is therefore governed by four matrix-matrix multiplications. With an additional multiplication with ${\bf S}^{-1}$ to form $\delta {\bf D}$,
which is not required in the DFTB example below, the additional cost of an extra rank update in Algorithm 3 includes five matrix-matrix multiplications.
This can be compared to the cost of a diagonalization required in each additional SCF iteration of a regular direct Born-Oppenheimer molecular dynamics scheme.

\linespread{1}
\begin{algorithm}[H]
\label{XLBOMD}
\caption{Pseudocode for the density-matrix based  XL-BOMD simulation scheme (after an initial ground state optimization).}
\algsetup{indent=1em}
\begin{algorithmic}
\STATE $ \delta t, ~ N_{\rm steps} ~~ [\mbox{Size and number of integration time steps}]$
\STATE $ {\bf R}(t_0) = {\bf R}_0, ~ {\bf {\dot R}}(t_0) = {\bf{\dot  R}}_0 ~~ [\mbox{Coordinates and velocities at $t = t_0$}]$
\STATE $ {\bf D} = {\bf D}_{\rm min} ~~ [\mbox{Initial relaxed ground state density matrix for ${\bf R}_0$ at $t = t_0$}]$
\STATE $ {\bf S} = {\bf S}[{\bf R}(t_0)], ~ {\bf Z} = {\bf S}^{-1/2} ~~ [\mbox{Initial overlap matrix and inverse overlap factor at $t = t_0$}]$
\STATE $ {\cal K}_0 = {\cal I} ~~ [\mbox{No preconditioner used here}]$
\STATE $ {\bf X}(t_0-k\delta t) = {\bf D}{\bf S}, ~~ k = 0,1,2,\ldots,k_{\rm max}$
\STATE $ c_k, ~ \alpha, ~ \kappa = \delta t^2 \omega^2 ~~ [\mbox{Integration coefficients for the Verlet damping term, Ref.\ \cite{ANiklasson09}}]$
\STATE $ {\cal U}({\bf R},{\bf X}) = U({\bf R}_0)~~ [\mbox{Initial potential energy at $t = t_0$}]$
\STATE $ \partial {\cal U}({\bf R},{\bf X})/\partial R_I = \partial U({\bf R}_0)/\partial R_I ~~[\mbox{Initial force at $t=t_0$}]$
\FOR{i = 1, $N_{\rm steps}$}
 \STATE $ t = t_0 + (i-1)\delta t$
 \STATE $ {\rm E}_{\rm tot}(t) = (1/2)\sum_I M_I {\dot R}_I^2 + {\cal U}({\bf R},{\bf X})$ ~~[Constant of motion]
 \STATE $ {\dot R}_I(t+\delta t/2) = {\dot R}_I(t)  - (\delta t/2)M_I^{-1}\partial{\cal U}({\bf R},{\bf X})/\partial R_I$ ~~[first half of leapfrog step]
 \STATE $ {\bf W}_0 = {\cal K}_0({\bf D}{\bf S}-{\bf X}) ~~ [\mbox{Residual}]$  
  \IF { i = 1}
   \STATE ${\ddot {\bf X}}(t) = {\bf W}_0$
  \ELSE
    \STATE $ m = 0$
    \WHILE {$({\rm Error} > {\rm Error~ Tolerance})$}
      \STATE $ m = m + 1$
      \STATE $ {\bf V}_m  = {\bf W}_{m-1}$
      \FOR{j = 1, m-1}
         \STATE ${\bf V}_m = {\bf V}_m - \langle {\bf V}_m, {\bf V}_j \rangle {\bf V}_j$ ~~[\mbox{Orthogonalization}]
      \ENDFOR
      \STATE $ {\bf V}_m = {\bf V}_m/\| {\bf V}_m\|, ~~ \delta {\bf D} = {\bf V}_m {\bf S}^{-1}$
      \STATE $ {\bf F}_1^\perp = {\bf Z}^T{\bf G}[{\bf R}(t),\delta {\bf D}] {\bf Z}$
      \STATE $ {\bf D}_1^\perp = \partial {\bf D}^\perp[{\bf F}_0^\perp + \lambda {\bf F}_1^\perp]/\partial \lambda~ \vert_{\lambda = 0}$ ~~ [use Alg.\ 2]
      \STATE $ {\bf W}_m = {\cal K}_0({\bf Z} {\bf D}_1^\perp {\bf Z}^{-1} - {\bf V}_m) $
      \STATE $ O_{kl} = \langle {\bf W}_k, {\bf W}_l \rangle , ~~ k,l = 1,2,\ldots , m$
      \STATE $ {\cal I}_{Res}{\bf W}_0 = \sum_{k,l = 1}^m {\bf W}_k M_{kl} \langle {\bf W}_l, {\bf W}_0 \rangle, ~~ {\bf M} = {\bf O}^{-1} $
      \STATE $ {\rm Error} = \|{\cal I}_{Res}{\bf W}_0 -{\bf W}_0\|/\|{\bf W}_0\| $
    \ENDWHILE
    \STATE $ {\ddot {\bf X}}(t) = -\sum_{k,l = 1}^{m} {\bf V}_k M_{kl} \langle {\bf W}_l, {\bf W}_0 \rangle  $ ~~[as in Eqs.\ \ref{KernelAppr} and \ref{Xddot} ]
  \ENDIF
  \STATE $ {\bf X}(t+\delta t) = 2 {\bf X}(t) - {\bf X}(t-\delta t) + \kappa {\ddot {\bf X}}(t) + \alpha \sum_{k = 0}^{k_{\rm max}} c_k {\bf X}(t-k\delta t) $ ~~[as in Eq.\ \ref{IntEqX}]
  \STATE $ {R}_I(t+\delta t) = {R}_I(t)  - \delta t {\dot R}_I(t+\delta t/2)$ ~~[full leapfrog integration time step]
  \STATE $ {\bf S} = {\bf S}({\bf R}(t+\delta t)), ~~ {\bf Z} = {\bf S}^{-1/2}, ~~ {\bf P} = {\bf X}(t+\delta t){\bf S}^{-1}, ~~ V_{nn} = V_{nn}({\bf R}(t+\delta t))$ 
  \STATE $ {\bf G} = {\bf G}({\bf R}(t+\delta t),{\bf P}), ~~{\bf F} = {\bf h}({\bf R}(t+\delta t)) + {\bf G}, ~~ {\bf F}_0^\perp = {\bf Z}^T{\bf F}{\bf Z}$
  \STATE $ {\rm Diagonalize:}~ {\bf Q}^T{\bf F}_0^\perp {\bf Q} = {\bf E}, ~{\rm where}~ E_{ij} = \delta_{ij} \epsilon_i ~ {\rm and}~ {\bf Q}^T{\bf Q} = \delta_{ij}, ~~ f_i = \left(e^{\beta(\epsilon_i -\mu_0 )} + 1\right)^{-1}$
  \STATE $ {\bf D}^\perp = \left(e^{\beta({\bf F}_0^\perp -\mu_0 {\bf I})} + {\bf I}\right)^{-1} = {\bf Q}\left(e^{\beta({\bf E} -\mu_0 {\bf I})} + {\bf I}\right)^{-1}{\bf Q}^T, ~~ {\bf D} = {\bf Z}{\bf D}^\perp {\bf Z}^T $
  \STATE $ {\cal U}({\bf R}(t+\delta t),{\bf X}) = 2 {\rm Tr}[{\bf h}{\bf D}] + {\rm Tr}[(2{\bf D}-{\bf P}){\bf G}] -2 T_e {\cal S}({\bf f}) + V_{nn}$ ~~[ Eq.\ \ref{ShadowPot}]
  \STATE $ \partial {\cal U}({\bf R},{\bf X})/\partial R_I = 2{\rm Tr}[{\bf h}_{R_I} {\bf D}] + {\rm Tr}[(2{\bf D}-{\bf P}){\bf G}_{R_I}] + \partial V_{nn}/\partial R_I - 2 {\rm Tr} [{\bf Z}{\bf Z}^T{\bf F}{\bf D} {\bf S}_{R_I}] $
  \STATE $  {\dot R}_I(t+\delta t) = {\dot R}_I(t+\delta t/2)  - (\delta t/2)M_I^{-1}\partial {\cal U}({\bf R},{\bf X})/\partial R_I$ ~~ [second half of leapfrog step]
\ENDFOR
\end{algorithmic}
\end{algorithm}


\section{Examples}

To illustrate XL-BOMD for a density matrix formulation using the thermal restricted Hartree-Fock theory, I will first use
the example of a single hydrogen molecule with a basis set of 8 Gaussian atomic orbitals \cite{JMThijssen99}.
Thereafter, I will demonstrate a XL-BOMD simulations for a reactive mixture of nitromethane using semiempirical self-consistent charge 
density functional tight-binding theory (SCC-DFTB) \cite{MElstner98,MFinnis98,TFrauenheim00,MGaus11,BAradi15,BHourahine20}
as implemented in a developers version of the open-source electronic structure software package LATTE \cite{LATTE,MCawkwell12},
with the parametrization given in Ref.\ \cite{AKrishnapriyan17}.
The SCC-DFTB LATTE code was modified such that the density matrix was used as the extended electronic degrees of
freedom instead of the net Mulliken charges.
The implementations closely follow the general scheme in Algorithm 3 for the simulation based on both Hartree-Fock 
and SCC-DFTB theory. For the coefficients of the modified Verlet integration scheme of the electronic degrees
of freedom, $\{c_k\}$, $\alpha$ and $\kappa = \delta t^2 \omega^2$, 
I used the optimized values in Ref.\ \cite{ANiklasson09} with $k_{\rm max}= 6$ for the Hartree-Fock example
and with $k_{\rm max}= 5$ for the SCC-DFTB simulation. The number of recursion steps, $n$, used in the 
canonical density matrix perturbation scheme, Algorithm 2, was set to $n = 8$.

\begin{figure}
\includegraphics[scale=0.45]{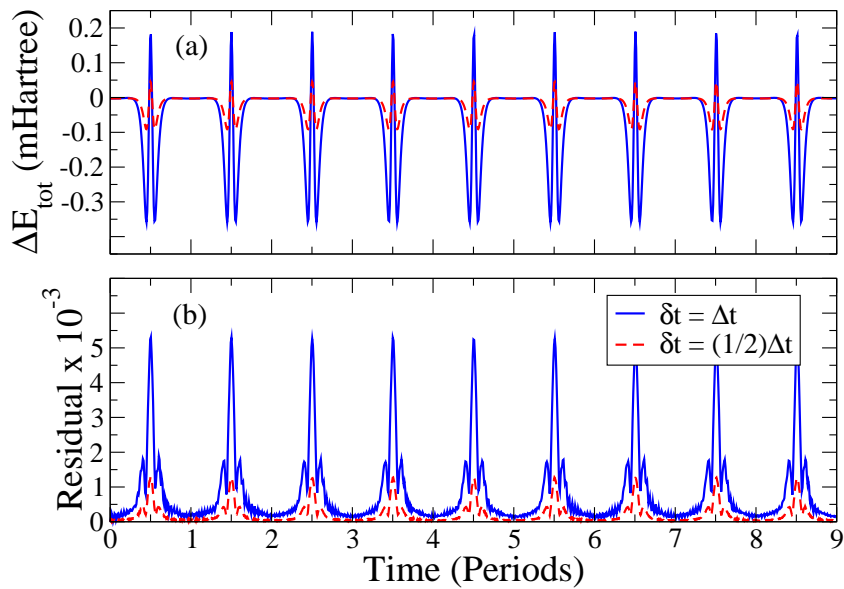}
\caption{\label{Fig_1}
{(a) Fluctuations in the total free energy, $\Delta E_{\rm tot}(t) = E_{\rm tot}(t)-E_{\rm tot}(t_0)$, 
where $E_{\rm tot} \equiv {\cal H}_{\rm XL-BOMD}$ in Eq.\ \ref{ConstH}, over nine periods of oscillations
for XL-BOMD simulations of a hydrogen molecule at $T_e = 1500$ K, using restricted, thermal Hartree-Fock theory in a density matrix formulation.
(b) Frobenius norm of 
the residual function, $\|({\bf D}[{\bf X}]{\bf S}-{\bf X})\|$, as a function of the period of the oscillations. The size of the fluctuations
approximately scales as $\sim \delta t^2$. The simulations were performed following the prototype scheme as outlined in the pseudocode
in Algorithm 3.}}
\end{figure}

\subsection{A Hydrogen Molecule Using Thermal Hartree-Fock Theory}

\begin{figure}
\includegraphics[scale=0.45]{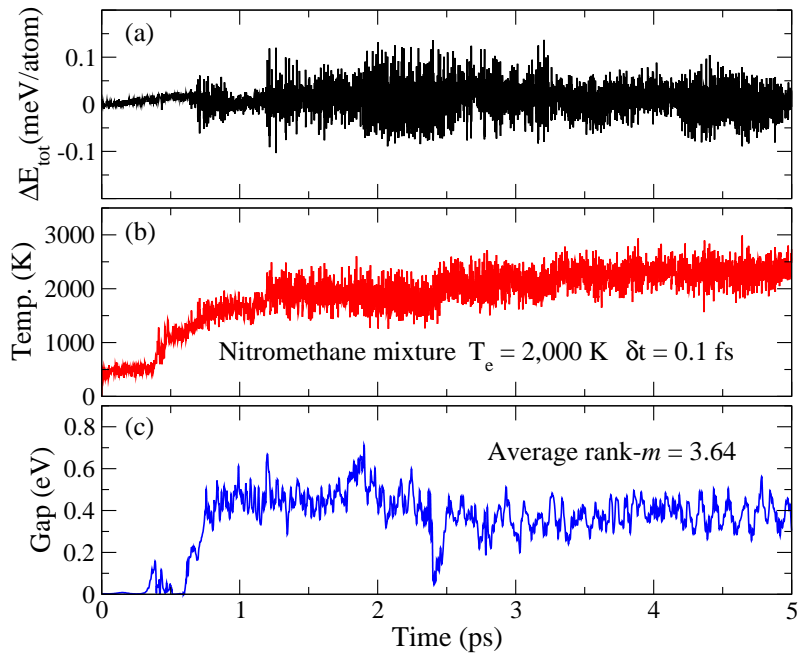}
\caption{\label{Fig_2}
{\small Results of an XL-BOMD simulation based on SCC-DFTB theory of a reactive mixture of liquid nitromethane, 
where the density matrix is used as the extended electronic degrees of freedom. 
The upper panel (a) shows the fluctuations in the total energy per atom, panel (b) shows the statistical
temperature (Temp.), panel (c) shows the electronic gap, $\epsilon_{N_{\rm occ}+1} - \epsilon_{N_{\rm occ}}$,
corresponding to the HOMO-LUMO gap at zero electronic temperature, $T_e = 0$.
The simulation was performed following the prototype pseudocode as outlined in Algorithm 3.}}
\end{figure}

The upper panel (a) of Figure \ref{Fig_2} shows the fluctuations in the total free energy, $\Delta E_{\rm tot}(t) = E_{\rm tot}(t)-E_{\rm tot}(t_0)$,
where $E_{\rm tot} \equiv {\cal H}_{\rm XL-BOMD}$ in Eq.\ \ref{ConstH}, for XL-BOMD simulations of a Hydrogen molecule at $T_e = 1500$ K.
The fluctuations correspond to the local truncation error of the leapfrog velocity Verlet integration 
of the nuclear and electronic degrees of freedom in Eq.\ \ref{Integration}.
The molecule is released without any initial velocity at an interatomic distance of 3 Bohr radius, which
is over twice the equilibrium distance. The trajectories were followed back and forth over 9 periods of oscillations.  
The lower panel (b) shows the Frobenius norm of the residual function, $\|({\bf D}[{\bf X}]{\bf S}-{\bf X})\|$, as a function of time. 
The figure shows the result of two simulations using two different integration time steps, $\delta t = \Delta t$ and $\delta t = \Delta t/2$.
As the integration time step is reduced by a factor of 2, the size of the fluctuations in the total energy and in
the residual function are reduced by a factor of about 4. This is what is expected from the
Verlet-based integration scheme and the scaling of the amplitudes serves as a sensitive gauge on 
the accuracy of the implementation together
with the absence of any visible systematic long-term drift in the total energy. The adaptive rank-$m$ update was used
together with a chosen relative error tolerance of $0.1$ for the error term (Error) in Algorithm 1. 
In all the simulations the rank, $m$, of the adaptive low-rank approximation was small, 
$m \le 3$, with an average of 2.25 for $\delta t = \Delta t$ and 2.27 for $\delta t = \Delta t/2$. This indicates that
the low-rank approximation of the kernel acting on the residual in the electronic degrees of freedom
can be approximated accurately at low cost, without the need for preconditioning.

\subsection{Reactive Nitromehtane Using Density Functional Tight-Binding Theory}

Figure \ref{Fig_2} shows the total energy fluctuations in the upper panel (a) of a reactive mixture of liquid nitromethane, with a few
randomly swapped positions between N, C, O, and H atoms. Here $\Delta E_{\rm tot}(t) = E_{\rm tot}(t)-E_{\rm tot}(t_0)$,
where $E_{\rm tot} \equiv {\cal H}_{\rm XL-BOMD}$ in Eq.\ \ref{ConstH}. 
This is a highly reactive molecular soup that most probably is unphysical, but it is a challenging simulation example and
therefore well suited to evaluate the density-matrix based formulation of XL-BOMD.
The statistical temperature (Temp.) quickly reaches over $2000$ K. 
At the same time the minute electronic HOMO-LUMO gap is rapidly opening and closing during the first
half picosecond of simulation time. Several exothermic reactions occur, seen as steps in the fluctuations of the statistical temperature
in the middle panel (b). After some of these reactions, the amplitude of the total energy fluctuations increases fast, but is then followed by a slower reduction. 
This behavior indicates rapid local reactions, creating a few fast moving molecules, which then is followed by a thermal equlibration.
The adaptive rank-$m$ update was used together with a chosen relative error tolerance of $0.1$ for the relative error term (Error) in Algorithm 1. 
In the simulation $m$ remained small, $m \le 5$, with an average ${\bar m}= 3.42$. 
This demonstrates that the low-rank approximation of the kernel works well 
without preconditioning also for this more challenging example compared to the hydrogen molecule.
The electronic temperature was set to $T_e = 2000$ K. The fractional occupation numbers are of importance to avoid instabilities
and too rapid changes in the electronic structure between time steps. The inclusion of an electronic temperature and entropy
reduces transition barriers.  A more precise {\em ab initio} description of bond breaking and bond formation may require 
multideterminant wave functions and nonadiabatic effects, which is beyond the scope of this article.

\section{Summary and Discussion}

I have presented an adaptive integration scheme for the extended electronic degrees of freedom in XL-BOMD simulations where
a density matrix with fractional occupation numbers is used as a dynamical tensor variable. This allows
sampling over regions of the energy landscape where the gap is small or vanishing, which leads
to particular convergence problems in regular direct Born-Oppenheimer molecular dynamics simulations.
The formulation and algorithms in this article were presented mainly in the language of Hartree-Fock theory, which is suitable
for quantum chemistry methods. However, the theory provides a general guide to implement XL-BOMD in a broad variety of electronic structure software
packages using a density matrix formalism
\cite{gamess,Gaussian94A,JStewart96,JSoler02,DBowler06,NHine09,LDMPeters17,RKendall00,JVandevondele05,BAradi07,ERudberg_11,DOseiKuffuor14,SVLevchenko15,BHourahine20}.
The main purpose of this article is to facilitate such implementations.

The density matrix formalism requires a fourth-order metric tensor, ${\cal T} \equiv {\cal K}^T{\cal K}$, 
in the generalized extended harmonic oscillator. The corresponding kernel, ${\cal K}$,
in the equations of motion for the extended electronic degrees of freedom then also represents a
fourth-order tensor mapping between matrices.
A tunable low-rank, Krylov subspace approximation of the kernel ${\cal K}$, which is defined as the inverse Jacobian of a residual matrix 
function, was presented for the integration of the extended electronic degrees of freedom. A computational scheme for the density-matrix based XL-BOMD 
simulations was given in pseudocode in Algorithm 3, which summarizes the main results of this article.
Examples for a single hydrogen molecule
and a reactive mixture of liquid nitromethane were used to demonstrate the prototype code. No iterative self-consistent field optimization
was required prior to the force evaluations (apart from the first time step), in contrast to regular direct 
Born-Oppenheimer molecular dynamics simulations. The tunable low-rank approximation worked well even without 
any preconditioner.

The matrix residual ${\bf D}{\bf S}-{\bf X}$ is a key quantity, which determines the size of the electronic force and provides a measure of the disctance to the
exact Born-Oppenheimer ground state. Its size scales quadratically with the integration time step, i.e.\ as $\delta t^2$. 
During reactions when the gap is rapidly dimished for a short period of time, 
the residual often goes up in size, but is then soon reduced. This dissipation is partly driven by the damping term in the 
modified Verlet integration of the extended electronic degrees of freedom, Eq.\ \ref{Integration}. The matrix residual can
be used to gauge the quality of the simulation and could possibly be used to adapt the time step.

The tunable rank-$m$ approximation is not fail-safe. It has the same limitations as Newton's method, which requires the initial guess
to be sufficiently close to the exact solution.  If the change in the electronic structure between time step is too large, 
the rank-$m$ approximation could fail and lead to instability and divergence. 

XL-BOMD may appear as a framework where we switch the problem of
the nonlinear self-consistent field optimization in direct Born-Oppenheimer molecular dynamics
to a kernel approximation that requires the solution of a system of nonlinear equations in the
integration of the extended electronic degrees of freedom. However, the kernel approximation is performed in
a single step, requiring no additional diagonalization, whereas a nonlinear self-consistent field optimization
requires multiple steps involving a diagonalization in each iteration. In the present (non-optimized) DFTB implementation used 
for the simulation of the reactive nitromethane mixture, the computational cost of each extra rank is dominated by the response calculation and its
required transformations into the molecular eigenbasis and then back to the atomic-orbital representation. This involves 4 matrix-matrix
multiplications and costs about half as much as the diagonalization, which is the additional cost in an extra SCF iteration.
For larger systems the overhead for an extra rank in the kernel approximation would go 
down as the relative cost of the diagonalization is likely to go up.
Moreover, the forces and the molecular trajectories often show little sensitivity to the accuracy of 
the kernel approximation. In fact, for nonreactive chemical systems, it is often sufficient to approximate
the kernel with a simple fixed scaled identity matrix. 
In contrast, the forces are not conservative in regular Born-Oppenheimer molecular dynamics,
unless a tight self-consistent field convergence is reached, and the system often
exhibits a systematic drift in the total energy where the system is artificially heated up or 
cooled down \cite{DRemler90}. The problems are particularly difficult for reactive chemical system
where it is often difficult and expensive to reach a sufficiently tight ground-state convergence 
in regular Born-Oppenheimer molecular dynamics simultions when the gap is small or vanishing.
Moreover, to use restarts from overlapping atomic densities in each new time step to avoid
the systematic drift by restoring the time-reversal symmetry, further increases the computational cost 
and may introduce discontinuities if the solutions converge to different local minimima between time steps.
These shortcomings cannot be avoided by including thermostats \cite{EMartinez15}.
In contrast, XL-BOMD combines computational speed with both stability and physical accuracy.

No specific preconditioner, ${\cal K}_0$, was proposed or used in the simulation examples. For the density-matrix based XL-BOMD,
the preconditioner cannot be constructed explicitly and its form would depend on the specific electronic structure method that is used. 
Some possibilities for the design of preconditioners would include, for example, preconditioners based on 
the response of the individual diagonal atomic blocks 
of the density matrix at some given structure or some other selected coarse grained partial response, 
an approximate kernel for a related model problem like in Kerker mixing \cite{GPKerker81},
or a fixed low-rank preconditioner based on individual molecules or fragments for some approximate structure. 
The use of a preconditionier is ideal in XL-BOMD if it can be reused over many time steps. 
Chemical reactions are often rare events separated by thousands of time steps and an update 
of the perconditioner would then add little to the total cost of a simulation.

XL-BOMD has many interesting similarities to extended Lagrangian Car-Parrinello molecular dynamics 
\cite{RCar85,DRemler90,GPastore91,FBornemann98,DMarx00,JHutter12} as is discussed in ref \cite{ANiklasson17}.
It also exhibits features similar to a new generation molecular dynamics schemes, where the electronic degrees
of freedom is extrapolated or propagated dynamically \cite{PPulay04,JMHerbert05,TDKuhne07,TAtsumi08,JFang16,ACoretti18,SBonella20},
as in Car-Parrinello molecular dynamics. In this way XL-BOMD can be viewed as a next generation extended Lagrangian 
first principles molecular dynamics.

The introduction of extended Lagrangian first principles molecular dynamics by Car and Parrinello in 1985 provided a 
first practical approach to general quantum-based molecular dynamics simulations at a theory level with predictive accuracy. 
Their innovation pioneered the rapid development of a number of advanced techniques, including
highly efficient methods for direct Born-Oppenheimer molecular dynamics simulations \cite{MCPayne92,TAArias_92,RBarnett93,GKresse93}
that currently dominate the field. However, the extended Lagrangian approach to first
principles molecular dynamics simulations may now, once again, appear to be the more efficient alternative.

\section{Acknowledgements}

This work is supported by the U.S. Department of Energy Office of Basic Energy Sciences (FWP LANLE8AN)
and by the U.S. Department of Energy through the Los Alamos National Laboratory.
Los Alamos National Laboratory is operated by Triad National Security, LLC, for the National Nuclear Security
Administration of the U.S. Department of Energy Contract No. 892333218NCA000001.
Discussions with Christian Negre, Marc Cawkwell, and Yu Zhang as well as generous contributions 
from Travis Peery at the T-Division international 10 bar Caf\'{e} are gratefully acknowledged.

\bibliography{mndo_new_x}

\begin{thebibliography}{100}%
\makeatletter
\providecommand \@ifxundefined [1]{%
 \@ifx{#1\undefined}
}%
\providecommand \@ifnum [1]{%
 \ifnum #1\expandafter \@firstoftwo
 \else \expandafter \@secondoftwo
 \fi
}%
\providecommand \@ifx [1]{%
 \ifx #1\expandafter \@firstoftwo
 \else \expandafter \@secondoftwo
 \fi
}%
\providecommand \natexlab [1]{#1}%
\providecommand \enquote  [1]{``#1''}%
\providecommand \bibnamefont  [1]{#1}%
\providecommand \bibfnamefont [1]{#1}%
\providecommand \citenamefont [1]{#1}%
\providecommand \href@noop [0]{\@secondoftwo}%
\providecommand \href [0]{\begingroup \@sanitize@url \@href}%
\providecommand \@href[1]{\@@startlink{#1}\@@href}%
\providecommand \@@href[1]{\endgroup#1\@@endlink}%
\providecommand \@sanitize@url [0]{\catcode `\\12\catcode `\$12\catcode
  `\&12\catcode `\#12\catcode `\^12\catcode `\_12\catcode `\%12\relax}%
\providecommand \@@startlink[1]{}%
\providecommand \@@endlink[0]{}%
\providecommand \url  [0]{\begingroup\@sanitize@url \@url }%
\providecommand \@url [1]{\endgroup\@href {#1}{\urlprefix }}%
\providecommand \urlprefix  [0]{URL }%
\providecommand \Eprint [0]{\href }%
\providecommand \doibase [0]{http://dx.doi.org/}%
\providecommand \selectlanguage [0]{\@gobble}%
\providecommand \bibinfo  [0]{\@secondoftwo}%
\providecommand \bibfield  [0]{\@secondoftwo}%
\providecommand \translation [1]{[#1]}%
\providecommand \BibitemOpen [0]{}%
\providecommand \bibitemStop [0]{}%
\providecommand \bibitemNoStop [0]{.\EOS\space}%
\providecommand \EOS [0]{\spacefactor3000\relax}%
\providecommand \BibitemShut  [1]{\csname bibitem#1\endcsname}%
\let\auto@bib@innerbib\@empty
\bibitem [{\citenamefont {Niklasson}(2008{\natexlab{a}})}]{ANiklasson08}%
  \BibitemOpen
  \bibfield  {author} {\bibinfo {author} {\bibfnamefont {A.~M.~N.}\
  \bibnamefont {Niklasson}},\ }\href@noop {} {\bibfield  {journal} {\bibinfo
  {journal} {Phys. Rev. Lett.}\ }\textbf {\bibinfo {volume} {100}},\ \bibinfo
  {pages} {123004} (\bibinfo {year} {2008}{\natexlab{a}})}\BibitemShut
  {NoStop}%
\bibitem [{\citenamefont {Steneteg}\ \emph {et~al.}(2010)\citenamefont
  {Steneteg}, \citenamefont {Abrikosov}, \citenamefont {Weber},\ and\
  \citenamefont {Niklasson}}]{PSteneteg10}%
  \BibitemOpen
  \bibfield  {author} {\bibinfo {author} {\bibfnamefont {P.}~\bibnamefont
  {Steneteg}}, \bibinfo {author} {\bibfnamefont {I.~A.}\ \bibnamefont
  {Abrikosov}}, \bibinfo {author} {\bibfnamefont {V.}~\bibnamefont {Weber}}, \
  and\ \bibinfo {author} {\bibfnamefont {A.~M.~N.}\ \bibnamefont {Niklasson}},\
  }\href@noop {} {\bibfield  {journal} {\bibinfo  {journal} {Phys. Rev. B}\
  }\textbf {\bibinfo {volume} {82}},\ \bibinfo {pages} {075110} (\bibinfo
  {year} {2010})}\BibitemShut {NoStop}%
\bibitem [{\citenamefont {Zheng}\ \emph {et~al.}(2011)\citenamefont {Zheng},
  \citenamefont {Niklasson},\ and\ \citenamefont {Karplus}}]{GZheng11}%
  \BibitemOpen
  \bibfield  {author} {\bibinfo {author} {\bibfnamefont {G.}~\bibnamefont
  {Zheng}}, \bibinfo {author} {\bibfnamefont {A.~M.~N.}\ \bibnamefont
  {Niklasson}}, \ and\ \bibinfo {author} {\bibfnamefont {M.}~\bibnamefont
  {Karplus}},\ }\href@noop {} {\bibfield  {journal} {\bibinfo  {journal} {J.
  Chem. Phys.}\ }\textbf {\bibinfo {volume} {135}},\ \bibinfo {pages} {044122}
  (\bibinfo {year} {2011})}\BibitemShut {NoStop}%
\bibitem [{\citenamefont {Cawkwell}\ and\ \citenamefont
  {Niklasson}(2012)}]{MCawkwell12}%
  \BibitemOpen
  \bibfield  {author} {\bibinfo {author} {\bibfnamefont {M.~J.}\ \bibnamefont
  {Cawkwell}}\ and\ \bibinfo {author} {\bibfnamefont {A.~M.~N.}\ \bibnamefont
  {Niklasson}},\ }\href@noop {} {\bibfield  {journal} {\bibinfo  {journal} {J.
  Chem. Phys.}\ }\textbf {\bibinfo {volume} {137}},\ \bibinfo {pages} {134105}
  (\bibinfo {year} {2012})}\BibitemShut {NoStop}%
\bibitem [{\citenamefont {Hutter}(2012)}]{JHutter12}%
  \BibitemOpen
  \bibfield  {author} {\bibinfo {author} {\bibfnamefont {J.}~\bibnamefont
  {Hutter}},\ }\href@noop {} {\bibfield  {journal} {\bibinfo  {journal} {WIREs
  Comput. Mol. Sci.}\ }\textbf {\bibinfo {volume} {2}},\ \bibinfo {pages} {604}
  (\bibinfo {year} {2012})}\BibitemShut {NoStop}%
\bibitem [{\citenamefont {Lin}\ \emph {et~al.}(2014)\citenamefont {Lin},
  \citenamefont {Lu},\ and\ \citenamefont {Shao}}]{LLin14}%
  \BibitemOpen
  \bibfield  {author} {\bibinfo {author} {\bibfnamefont {L.}~\bibnamefont
  {Lin}}, \bibinfo {author} {\bibfnamefont {J.}~\bibnamefont {Lu}}, \ and\
  \bibinfo {author} {\bibfnamefont {S.}~\bibnamefont {Shao}},\ }\href@noop {}
  {\bibfield  {journal} {\bibinfo  {journal} {Entropy}\ }\textbf {\bibinfo
  {volume} {16}},\ \bibinfo {pages} {110} (\bibinfo {year} {2014})}\BibitemShut
  {NoStop}%
\bibitem [{\citenamefont {Arita}\ \emph {et~al.}(2014)\citenamefont {Arita},
  \citenamefont {Bowler},\ and\ \citenamefont {Miyazaki}}]{MArita14}%
  \BibitemOpen
  \bibfield  {author} {\bibinfo {author} {\bibfnamefont {M.}~\bibnamefont
  {Arita}}, \bibinfo {author} {\bibfnamefont {D.~R.}\ \bibnamefont {Bowler}}, \
  and\ \bibinfo {author} {\bibfnamefont {T.}~\bibnamefont {Miyazaki}},\
  }\href@noop {} {\bibfield  {journal} {\bibinfo  {journal} {J. Chem. Theory
  Comput.}\ }\textbf {\bibinfo {volume} {10}},\ \bibinfo {pages} {5419}
  (\bibinfo {year} {2014})}\BibitemShut {NoStop}%
\bibitem [{\citenamefont {Souvatzis}\ and\ \citenamefont
  {Niklasson}(2014)}]{PSouvatzis14}%
  \BibitemOpen
  \bibfield  {author} {\bibinfo {author} {\bibfnamefont {P.}~\bibnamefont
  {Souvatzis}}\ and\ \bibinfo {author} {\bibfnamefont {A.~M.~N.}\ \bibnamefont
  {Niklasson}},\ }\href@noop {} {\bibfield  {journal} {\bibinfo  {journal} {J.
  Chem. Phys.}\ }\textbf {\bibinfo {volume} {140}},\ \bibinfo {pages} {044117}
  (\bibinfo {year} {2014})}\BibitemShut {NoStop}%
\bibitem [{\citenamefont {Niklasson}\ and\ \citenamefont
  {Cawkwell}(2014)}]{ANiklasson14}%
  \BibitemOpen
  \bibfield  {author} {\bibinfo {author} {\bibfnamefont {A.~M.~N.}\
  \bibnamefont {Niklasson}}\ and\ \bibinfo {author} {\bibfnamefont
  {M.}~\bibnamefont {Cawkwell}},\ }\href@noop {} {\bibfield  {journal}
  {\bibinfo  {journal} {J. Chem. Phys.}\ }\textbf {\bibinfo {volume} {141}},\
  \bibinfo {pages} {164123} (\bibinfo {year} {2014})}\BibitemShut {NoStop}%
\bibitem [{\citenamefont {Nomura}\ \emph {et~al.}(2015)\citenamefont {Nomura},
  \citenamefont {Small}, \citenamefont {Kalia}, \citenamefont {Nakano},\ and\
  \citenamefont {Vashista}}]{KNomura15}%
  \BibitemOpen
  \bibfield  {author} {\bibinfo {author} {\bibfnamefont {K.}~\bibnamefont
  {Nomura}}, \bibinfo {author} {\bibfnamefont {P.~E.}\ \bibnamefont {Small}},
  \bibinfo {author} {\bibfnamefont {R.~K.}\ \bibnamefont {Kalia}}, \bibinfo
  {author} {\bibfnamefont {A.}~\bibnamefont {Nakano}}, \ and\ \bibinfo {author}
  {\bibfnamefont {P.}~\bibnamefont {Vashista}},\ }\href@noop {} {\bibfield
  {journal} {\bibinfo  {journal} {Comput. Phys. Comm.}\ }\textbf {\bibinfo
  {volume} {192}},\ \bibinfo {pages} {91} (\bibinfo {year} {2015})}\BibitemShut
  {NoStop}%
\bibitem [{\citenamefont {Albaugh}\ \emph {et~al.}(2015)\citenamefont
  {Albaugh}, \citenamefont {Demardash},\ and\ \citenamefont
  {Head-Gordon}}]{AAlbaugh15}%
  \BibitemOpen
  \bibfield  {author} {\bibinfo {author} {\bibfnamefont {A.}~\bibnamefont
  {Albaugh}}, \bibinfo {author} {\bibfnamefont {O.}~\bibnamefont {Demardash}},
  \ and\ \bibinfo {author} {\bibfnamefont {T.}~\bibnamefont {Head-Gordon}},\
  }\href@noop {} {\bibfield  {journal} {\bibinfo  {journal} {J. Chem. Phys.}\
  }\textbf {\bibinfo {volume} {143}},\ \bibinfo {pages} {174104} (\bibinfo
  {year} {2015})}\BibitemShut {NoStop}%
\bibitem [{\citenamefont {Negre}\ \emph {et~al.}(2016)\citenamefont {Negre},
  \citenamefont {Mnizsewski}, \citenamefont {Cawkwell}, \citenamefont {Bock},
  \citenamefont {Wall},\ and\ \citenamefont {Niklasson}}]{CNegre16}%
  \BibitemOpen
  \bibfield  {author} {\bibinfo {author} {\bibfnamefont {C.~F.~A.}\
  \bibnamefont {Negre}}, \bibinfo {author} {\bibfnamefont {S.~M.}\ \bibnamefont
  {Mnizsewski}}, \bibinfo {author} {\bibfnamefont {M.~J.}\ \bibnamefont
  {Cawkwell}}, \bibinfo {author} {\bibfnamefont {N.}~\bibnamefont {Bock}},
  \bibinfo {author} {\bibfnamefont {M.~E.}\ \bibnamefont {Wall}}, \ and\
  \bibinfo {author} {\bibfnamefont {A.~M.~N.}\ \bibnamefont {Niklasson}},\
  }\href@noop {} {\bibfield  {journal} {\bibinfo  {journal} {J. Chem. Theory
  Comput.}\ }\textbf {\bibinfo {volume} {12}},\ \bibinfo {pages} {3063}
  (\bibinfo {year} {2016})}\BibitemShut {NoStop}%
\bibitem [{\citenamefont {Niklasson}(2017)}]{ANiklasson17}%
  \BibitemOpen
  \bibfield  {author} {\bibinfo {author} {\bibfnamefont {A.~M.~N.}\
  \bibnamefont {Niklasson}},\ }\href@noop {} {\bibfield  {journal} {\bibinfo
  {journal} {J. Chem. Phys.}\ }\textbf {\bibinfo {volume} {147}},\ \bibinfo
  {pages} {054103} (\bibinfo {year} {2017})}\BibitemShut {NoStop}%
\bibitem [{\citenamefont {Bjorgaard}\ \emph {et~al.}(2018)\citenamefont
  {Bjorgaard}, \citenamefont {Sheppard}, \citenamefont {Tretiak},\ and\
  \citenamefont {Niklasson}}]{JBjorgaard18}%
  \BibitemOpen
  \bibfield  {author} {\bibinfo {author} {\bibfnamefont {J.~A.}\ \bibnamefont
  {Bjorgaard}}, \bibinfo {author} {\bibfnamefont {D.}~\bibnamefont {Sheppard}},
  \bibinfo {author} {\bibfnamefont {S.}~\bibnamefont {Tretiak}}, \ and\
  \bibinfo {author} {\bibfnamefont {A.~M.~N.}\ \bibnamefont {Niklasson}},\
  }\href {\doibase 10.1021/acs.jctc.7b00857} {\bibfield  {journal} {\bibinfo
  {journal} {Journal of Chemical Theory and Computation}\ }\textbf {\bibinfo
  {volume} {14}},\ \bibinfo {pages} {799} (\bibinfo {year} {2018})},\ \bibinfo
  {note} {pMID: 29316401},\ \Eprint
  {http://arxiv.org/abs/https://doi.org/10.1021/acs.jctc.7b00857}
  {https://doi.org/10.1021/acs.jctc.7b00857} \BibitemShut {NoStop}%
\bibitem [{\citenamefont {Car}\ and\ \citenamefont
  {Parrinello}(1985)}]{RCar85}%
  \BibitemOpen
  \bibfield  {author} {\bibinfo {author} {\bibfnamefont {R.}~\bibnamefont
  {Car}}\ and\ \bibinfo {author} {\bibfnamefont {M.}~\bibnamefont
  {Parrinello}},\ }\href@noop {} {\bibfield  {journal} {\bibinfo  {journal}
  {Phys. Rev. Lett.}\ }\textbf {\bibinfo {volume} {55}},\ \bibinfo {pages}
  {2471} (\bibinfo {year} {1985})}\BibitemShut {NoStop}%
\bibitem [{\citenamefont {Remler}\ and\ \citenamefont
  {Madden}(1990)}]{DRemler90}%
  \BibitemOpen
  \bibfield  {author} {\bibinfo {author} {\bibfnamefont {D.~K.}\ \bibnamefont
  {Remler}}\ and\ \bibinfo {author} {\bibfnamefont {P.~A.}\ \bibnamefont
  {Madden}},\ }\href@noop {} {\bibfield  {journal} {\bibinfo  {journal} {Mol.\
  Phys.}\ }\textbf {\bibinfo {volume} {70}},\ \bibinfo {pages} {921} (\bibinfo
  {year} {1990})}\BibitemShut {NoStop}%
\bibitem [{\citenamefont {Pastore}\ \emph {et~al.}(1991)\citenamefont
  {Pastore}, \citenamefont {Smargassi},\ and\ \citenamefont
  {Buda}}]{GPastore91}%
  \BibitemOpen
  \bibfield  {author} {\bibinfo {author} {\bibfnamefont {G.}~\bibnamefont
  {Pastore}}, \bibinfo {author} {\bibfnamefont {E.}~\bibnamefont {Smargassi}},
  \ and\ \bibinfo {author} {\bibfnamefont {F.}~\bibnamefont {Buda}},\
  }\href@noop {} {\bibfield  {journal} {\bibinfo  {journal} {Phys. Rev. A}\
  }\textbf {\bibinfo {volume} {44}},\ \bibinfo {pages} {6334} (\bibinfo {year}
  {1991})}\BibitemShut {NoStop}%
\bibitem [{\citenamefont {Bornemann}\ and\ \citenamefont
  {Sch\"{u}tte}(1998)}]{FBornemann98}%
  \BibitemOpen
  \bibfield  {author} {\bibinfo {author} {\bibfnamefont {F.~A.}\ \bibnamefont
  {Bornemann}}\ and\ \bibinfo {author} {\bibfnamefont {C.}~\bibnamefont
  {Sch\"{u}tte}},\ }\href@noop {} {\bibfield  {journal} {\bibinfo  {journal}
  {Numerische Mathematik}\ }\textbf {\bibinfo {volume} {78}},\ \bibinfo {pages}
  {359} (\bibinfo {year} {1998})}\BibitemShut {NoStop}%
\bibitem [{\citenamefont {Marx}\ and\ \citenamefont {Hutter}(2000)}]{DMarx00}%
  \BibitemOpen
  \bibfield  {author} {\bibinfo {author} {\bibfnamefont {D.}~\bibnamefont
  {Marx}}\ and\ \bibinfo {author} {\bibfnamefont {J.}~\bibnamefont {Hutter}},\
  }\enquote {\bibinfo {title} {Modern methods and algorithms of quantum
  chemistry},}\ \ (\bibinfo  {publisher} {ed. J. Grotendorst},\ \bibinfo
  {address} {John von Neumann Institute for Computing, J\"ulich, Germany},\
  \bibinfo {year} {2000})\ \bibinfo {edition} {2nd}\ ed.\BibitemShut {Stop}%
\bibitem [{\citenamefont {Niklasson}\ \emph {et~al.}(2011)\citenamefont
  {Niklasson}, \citenamefont {Steneteg},\ and\ \citenamefont
  {Bock}}]{ANiklasson11}%
  \BibitemOpen
  \bibfield  {author} {\bibinfo {author} {\bibfnamefont {A.~M.~N.}\
  \bibnamefont {Niklasson}}, \bibinfo {author} {\bibfnamefont {P.}~\bibnamefont
  {Steneteg}}, \ and\ \bibinfo {author} {\bibfnamefont {N.}~\bibnamefont
  {Bock}},\ }\href@noop {} {\bibfield  {journal} {\bibinfo  {journal} {J. Chem.
  Phys.}\ }\textbf {\bibinfo {volume} {135}},\ \bibinfo {pages} {164111}
  (\bibinfo {year} {2011})}\BibitemShut {NoStop}%
\bibitem [{\citenamefont {Souvatzis}\ and\ \citenamefont
  {Niklasson}(2013)}]{PSouvatzis13}%
  \BibitemOpen
  \bibfield  {author} {\bibinfo {author} {\bibfnamefont {P.}~\bibnamefont
  {Souvatzis}}\ and\ \bibinfo {author} {\bibfnamefont {A.~M.~N.}\ \bibnamefont
  {Niklasson}},\ }\href@noop {} {\bibfield  {journal} {\bibinfo  {journal} {J.
  Chem. Phys.}\ }\textbf {\bibinfo {volume} {139}},\ \bibinfo {pages} {214102}
  (\bibinfo {year} {2013})}\BibitemShut {NoStop}%
\bibitem [{\citenamefont {PetaChem}\ and\ \citenamefont
  {LLC}(2020)}]{TeraChem}%
  \BibitemOpen
  \bibfield  {author} {\bibinfo {author} {\bibnamefont {PetaChem}}\ and\
  \bibinfo {author} {\bibnamefont {LLC}},\ }\href@noop {} {\bibfield  {journal}
  {\bibinfo  {journal} {http://www.petachem.com/doc/userguide.pdf}\ } (\bibinfo
  {year} {2020})}\BibitemShut {NoStop}%
\bibitem [{\citenamefont {Aradi}\ \emph {et~al.}(2015)\citenamefont {Aradi},
  \citenamefont {Niklasson},\ and\ \citenamefont {Frauenheim}}]{BAradi15}%
  \BibitemOpen
  \bibfield  {author} {\bibinfo {author} {\bibfnamefont {B.}~\bibnamefont
  {Aradi}}, \bibinfo {author} {\bibfnamefont {A.~M.~N.}\ \bibnamefont
  {Niklasson}}, \ and\ \bibinfo {author} {\bibfnamefont {T.}~\bibnamefont
  {Frauenheim}},\ }\href@noop {} {\bibfield  {journal} {\bibinfo  {journal} {J.
  Chem. Theory Comput.}\ }\textbf {\bibinfo {volume} {11}},\ \bibinfo {pages}
  {3357} (\bibinfo {year} {2015})}\BibitemShut {NoStop}%
\bibitem [{\citenamefont {Vitale}\ \emph {et~al.}(2017)\citenamefont {Vitale},
  \citenamefont {Dziezic}, \citenamefont {Albaugh}, \citenamefont {Niklasson},
  \citenamefont {Head-Gordon},\ and\ \citenamefont {Skylaris}}]{VVitale17}%
  \BibitemOpen
  \bibfield  {author} {\bibinfo {author} {\bibfnamefont {V.}~\bibnamefont
  {Vitale}}, \bibinfo {author} {\bibfnamefont {J.}~\bibnamefont {Dziezic}},
  \bibinfo {author} {\bibfnamefont {A.}~\bibnamefont {Albaugh}}, \bibinfo
  {author} {\bibfnamefont {A.}~\bibnamefont {Niklasson}}, \bibinfo {author}
  {\bibfnamefont {T.~J.}\ \bibnamefont {Head-Gordon}}, \ and\ \bibinfo {author}
  {\bibfnamefont {C.-K.}\ \bibnamefont {Skylaris}},\ }\href@noop {} {\bibfield
  {journal} {\bibinfo  {journal} {J. Chem. Phys.}\ }\textbf {\bibinfo {volume}
  {12}},\ \bibinfo {pages} {124115} (\bibinfo {year} {2017})}\BibitemShut
  {NoStop}%
\bibitem [{\citenamefont {Lagardere}\ \emph {et~al.}(2018)\citenamefont
  {Lagardere}, \citenamefont {Jolly}, \citenamefont {Lipparini}, \citenamefont
  {Aviat}, \citenamefont {Stamm}, \citenamefont {Jing}, \citenamefont {Harger},
  \citenamefont {Torabifard}, \citenamefont {Cisneros}, \citenamefont
  {Schnieders}, \citenamefont {Gresh}, \citenamefont {Maday}, \citenamefont
  {Ren}, \citenamefont {Ponder},\ and\ \citenamefont {Piquemal}}]{Tinker-HP}%
  \BibitemOpen
  \bibfield  {author} {\bibinfo {author} {\bibfnamefont {L.}~\bibnamefont
  {Lagardere}}, \bibinfo {author} {\bibfnamefont {L.-H.}\ \bibnamefont
  {Jolly}}, \bibinfo {author} {\bibfnamefont {F.}~\bibnamefont {Lipparini}},
  \bibinfo {author} {\bibfnamefont {F.}~\bibnamefont {Aviat}}, \bibinfo
  {author} {\bibfnamefont {B.}~\bibnamefont {Stamm}}, \bibinfo {author}
  {\bibfnamefont {Z.~F.}\ \bibnamefont {Jing}}, \bibinfo {author}
  {\bibfnamefont {M.}~\bibnamefont {Harger}}, \bibinfo {author} {\bibfnamefont
  {H.}~\bibnamefont {Torabifard}}, \bibinfo {author} {\bibfnamefont {G.~A.}\
  \bibnamefont {Cisneros}}, \bibinfo {author} {\bibfnamefont {M.~J.}\
  \bibnamefont {Schnieders}}, \bibinfo {author} {\bibfnamefont
  {N.}~\bibnamefont {Gresh}}, \bibinfo {author} {\bibfnamefont
  {Y.}~\bibnamefont {Maday}}, \bibinfo {author} {\bibfnamefont {P.~Y.}\
  \bibnamefont {Ren}}, \bibinfo {author} {\bibfnamefont {J.~W.}\ \bibnamefont
  {Ponder}}, \ and\ \bibinfo {author} {\bibfnamefont {J.-P.}\ \bibnamefont
  {Piquemal}},\ }\href {\doibase 10.1039/C7SC04531J} {\bibfield  {journal}
  {\bibinfo  {journal} {Chem. Sci.}\ ,\ } (\bibinfo {year} {2018})}\BibitemShut
  {NoStop}%
\bibitem [{\citenamefont {Peters}\ \emph {et~al.}(2015)\citenamefont {Peters},
  \citenamefont {Kussmann},\ and\ \citenamefont {Ochsenfeld}}]{LDMPeters17}%
  \BibitemOpen
  \bibfield  {author} {\bibinfo {author} {\bibfnamefont {L.~D.~M.}\
  \bibnamefont {Peters}}, \bibinfo {author} {\bibfnamefont {J.}~\bibnamefont
  {Kussmann}}, \ and\ \bibinfo {author} {\bibfnamefont {C.}~\bibnamefont
  {Ochsenfeld}},\ }\href@noop {} {\bibfield  {journal} {\bibinfo  {journal} {J.
  Chem. Theory Comput.}\ }\textbf {\bibinfo {volume} {13}},\ \bibinfo {pages}
  {5479} (\bibinfo {year} {2015})}\BibitemShut {NoStop}%
\bibitem [{\citenamefont {Otsuka}\ \emph {et~al.}(2016)\citenamefont {Otsuka},
  \citenamefont {Taiji}, \citenamefont {Bowler},\ and\ \citenamefont
  {Miyazaki}}]{TOtsuka16}%
  \BibitemOpen
  \bibfield  {author} {\bibinfo {author} {\bibfnamefont {T.}~\bibnamefont
  {Otsuka}}, \bibinfo {author} {\bibfnamefont {M.}~\bibnamefont {Taiji}},
  \bibinfo {author} {\bibfnamefont {D.~R.}\ \bibnamefont {Bowler}}, \ and\
  \bibinfo {author} {\bibfnamefont {T.}~\bibnamefont {Miyazaki}},\ }\href
  {http://stacks.iop.org/1347-4065/55/i=11/a=1102B1} {\bibfield  {journal}
  {\bibinfo  {journal} {Japanese Journal of Applied Physics}\ }\textbf
  {\bibinfo {volume} {55}},\ \bibinfo {pages} {1102B1} (\bibinfo {year}
  {2016})}\BibitemShut {NoStop}%
\bibitem [{\citenamefont {Hirakawa}\ \emph {et~al.}(2017)\citenamefont
  {Hirakawa}, \citenamefont {suzuki}, \citenamefont {Bowler},\ and\
  \citenamefont {Myazaki}}]{THirakawa17}%
  \BibitemOpen
  \bibfield  {author} {\bibinfo {author} {\bibfnamefont {T.}~\bibnamefont
  {Hirakawa}}, \bibinfo {author} {\bibfnamefont {T.}~\bibnamefont {suzuki}},
  \bibinfo {author} {\bibfnamefont {D.~R.}\ \bibnamefont {Bowler}}, \ and\
  \bibinfo {author} {\bibfnamefont {T.}~\bibnamefont {Myazaki}},\ }\href@noop
  {} {\bibfield  {journal} {\bibinfo  {journal} {J. Phys.: Condens. Matter}\
  }\textbf {\bibinfo {volume} {29}},\ \bibinfo {pages} {405901} (\bibinfo
  {year} {2017})}\BibitemShut {NoStop}%
\bibitem [{\citenamefont {Albaugh}\ \emph {et~al.}(2017)\citenamefont
  {Albaugh}, \citenamefont {Niklasson},\ and\ \citenamefont
  {Head-Gordon}}]{AAlbaugh17}%
  \BibitemOpen
  \bibfield  {author} {\bibinfo {author} {\bibfnamefont {A.}~\bibnamefont
  {Albaugh}}, \bibinfo {author} {\bibfnamefont {A.}~\bibnamefont {Niklasson}},
  \ and\ \bibinfo {author} {\bibfnamefont {T.}~\bibnamefont {Head-Gordon}},\
  }\href@noop {} {\bibfield  {journal} {\bibinfo  {journal} {J. Phys. Chem.
  Lett.}\ }\textbf {\bibinfo {volume} {8}},\ \bibinfo {pages} {1714} (\bibinfo
  {year} {2017})}\BibitemShut {NoStop}%
\bibitem [{\citenamefont {Albaugh}\ and\ \citenamefont
  {Head-Gordon}(2017)}]{AAlbaugh17b}%
  \BibitemOpen
  \bibfield  {author} {\bibinfo {author} {\bibfnamefont {A.}~\bibnamefont
  {Albaugh}}\ and\ \bibinfo {author} {\bibfnamefont {T.}~\bibnamefont
  {Head-Gordon}},\ }\href@noop {} {\bibfield  {journal} {\bibinfo  {journal}
  {J. Chem. Theory Comput.}\ }\textbf {\bibinfo {volume} {13}},\ \bibinfo
  {pages} {5207} (\bibinfo {year} {2017})}\BibitemShut {NoStop}%
\bibitem [{\citenamefont {Albaugh}\ \emph {et~al.}(2018)\citenamefont
  {Albaugh}, \citenamefont {Head-Gordon},\ and\ \citenamefont
  {Niklasson}}]{AAlbaugh18}%
  \BibitemOpen
  \bibfield  {author} {\bibinfo {author} {\bibfnamefont {A.}~\bibnamefont
  {Albaugh}}, \bibinfo {author} {\bibfnamefont {T.}~\bibnamefont
  {Head-Gordon}}, \ and\ \bibinfo {author} {\bibfnamefont {A.~M.~N.}\
  \bibnamefont {Niklasson}},\ }\href {\doibase 10.1021/acs.jctc.7b01041}
  {\bibfield  {journal} {\bibinfo  {journal} {Journal of Chemical Theory and
  Computation}\ }\textbf {\bibinfo {volume} {14}},\ \bibinfo {pages} {499}
  (\bibinfo {year} {2018})},\ \bibinfo {note} {pMID: 29316388},\ \Eprint
  {http://arxiv.org/abs/https://doi.org/10.1021/acs.jctc.7b01041}
  {https://doi.org/10.1021/acs.jctc.7b01041} \BibitemShut {NoStop}%
\bibitem [{\citenamefont {Henning}\ and\ \citenamefont
  {Niklasson}(2019)}]{PHenning19}%
  \BibitemOpen
  \bibfield  {author} {\bibinfo {author} {\bibfnamefont {P.}~\bibnamefont
  {Henning}}\ and\ \bibinfo {author} {\bibfnamefont {A.~M.~N.}\ \bibnamefont
  {Niklasson}},\ }\href@noop {} {\bibfield  {journal} {\bibinfo  {journal}
  {ArXive https://arxiv.org/abs/1912.10303}\ } (\bibinfo {year}
  {2019})}\BibitemShut {NoStop}%
\bibitem [{\citenamefont {Kroonblawd}\ \emph {et~al.}(2019)\citenamefont
  {Kroonblawd}, \citenamefont {Lindsey},\ and\ \citenamefont
  {Goldman}}]{MKroonblawd19}%
  \BibitemOpen
  \bibfield  {author} {\bibinfo {author} {\bibfnamefont {M.~P.}\ \bibnamefont
  {Kroonblawd}}, \bibinfo {author} {\bibfnamefont {R.~K.}\ \bibnamefont
  {Lindsey}}, \ and\ \bibinfo {author} {\bibfnamefont {N.}~\bibnamefont
  {Goldman}},\ }\href@noop {} {\bibfield  {journal} {\bibinfo  {journal} {Chem.
  Sci.}\ }\textbf {\bibinfo {volume} {10}},\ \bibinfo {pages} {6091} (\bibinfo
  {year} {2019})}\BibitemShut {NoStop}%
\bibitem [{\citenamefont {Kroonblawd}\ and\ \citenamefont
  {Goldman}(2020)}]{MKroonblawd20}%
  \BibitemOpen
  \bibfield  {author} {\bibinfo {author} {\bibfnamefont {M.~P.}\ \bibnamefont
  {Kroonblawd}}\ and\ \bibinfo {author} {\bibfnamefont {N.}~\bibnamefont
  {Goldman}},\ }\enquote {\bibinfo {title} {Free energies of reaction for
  aqueous glycine condensation chemistry at extreme temperatures},}\ in\
  \href@noop {} {\emph {\bibinfo {booktitle} {Carbon in Earth's Interior,
  Geophysical Monograph Series}}},\ \bibinfo {editor} {edited by\ \bibinfo
  {editor} {\bibfnamefont {C.~E.}\ \bibnamefont {Manning}}, \bibinfo {editor}
  {\bibfnamefont {J.}~\bibnamefont {Lin}}, \ and\ \bibinfo {editor}
  {\bibfnamefont {W.~L.}\ \bibnamefont {Mao}}}\ (\bibinfo  {publisher} {John
  Wiley and Sons, Inc.},\ \bibinfo {year} {2020})\ p.\ \bibinfo {pages}
  {271}\BibitemShut {NoStop}%
\bibitem [{\citenamefont {Niklasson}\ \emph {et~al.}(2006)\citenamefont
  {Niklasson}, \citenamefont {Tymczak},\ and\ \citenamefont
  {Challacombe}}]{ANiklasson06}%
  \BibitemOpen
  \bibfield  {author} {\bibinfo {author} {\bibfnamefont {A.~M.~N.}\
  \bibnamefont {Niklasson}}, \bibinfo {author} {\bibfnamefont {C.~J.}\
  \bibnamefont {Tymczak}}, \ and\ \bibinfo {author} {\bibfnamefont
  {M.}~\bibnamefont {Challacombe}},\ }\href@noop {} {\bibfield  {journal}
  {\bibinfo  {journal} {Phys. Rev. Lett.}\ }\textbf {\bibinfo {volume} {97}},\
  \bibinfo {pages} {123001} (\bibinfo {year} {2006})}\BibitemShut {NoStop}%
\bibitem [{\citenamefont {Niklasson}\ \emph {et~al.}(2009)\citenamefont
  {Niklasson}, \citenamefont {Steneteg}, \citenamefont {Odell}, \citenamefont
  {Bock}, \citenamefont {Challacombe}, \citenamefont {Tymczak}, \citenamefont
  {Holmstrom}, \citenamefont {Zheng},\ and\ \citenamefont
  {Weber}}]{ANiklasson09}%
  \BibitemOpen
  \bibfield  {author} {\bibinfo {author} {\bibfnamefont {A.~M.~N.}\
  \bibnamefont {Niklasson}}, \bibinfo {author} {\bibfnamefont {P.}~\bibnamefont
  {Steneteg}}, \bibinfo {author} {\bibfnamefont {A.}~\bibnamefont {Odell}},
  \bibinfo {author} {\bibfnamefont {N.}~\bibnamefont {Bock}}, \bibinfo {author}
  {\bibfnamefont {M.}~\bibnamefont {Challacombe}}, \bibinfo {author}
  {\bibfnamefont {C.~J.}\ \bibnamefont {Tymczak}}, \bibinfo {author}
  {\bibfnamefont {E.}~\bibnamefont {Holmstrom}}, \bibinfo {author}
  {\bibfnamefont {G.}~\bibnamefont {Zheng}}, \ and\ \bibinfo {author}
  {\bibfnamefont {V.}~\bibnamefont {Weber}},\ }\href@noop {} {\bibfield
  {journal} {\bibinfo  {journal} {J. Chem. Phys.}\ }\textbf {\bibinfo {volume}
  {130}},\ \bibinfo {pages} {214109} (\bibinfo {year} {2009})}\BibitemShut
  {NoStop}%
\bibitem [{\citenamefont {Niklasson}(2020{\natexlab{a}})}]{ANiklasson20}%
  \BibitemOpen
  \bibfield  {author} {\bibinfo {author} {\bibfnamefont {A.~M.~N.}\
  \bibnamefont {Niklasson}},\ }\href@noop {} {\bibfield  {journal} {\bibinfo
  {journal} {J. Chem. Phys.}\ }\textbf {\bibinfo {volume} {152}},\ \bibinfo
  {pages} {104103} (\bibinfo {year} {2020}{\natexlab{a}})}\BibitemShut
  {NoStop}%
\bibitem [{\citenamefont {Roothaan}(1951)}]{Roothaan}%
  \BibitemOpen
  \bibfield  {author} {\bibinfo {author} {\bibfnamefont {C.~C.~J.}\
  \bibnamefont {Roothaan}},\ }\href@noop {} {\bibfield  {journal} {\bibinfo
  {journal} {Rev. Mod. Phys.}\ }\textbf {\bibinfo {volume} {23}},\ \bibinfo
  {pages} {69} (\bibinfo {year} {1951})}\BibitemShut {NoStop}%
\bibitem [{\citenamefont {McWeeny}(1960)}]{RMcWeeny60}%
  \BibitemOpen
  \bibfield  {author} {\bibinfo {author} {\bibfnamefont {R.}~\bibnamefont
  {McWeeny}},\ }\href@noop {} {\bibfield  {journal} {\bibinfo  {journal} {Rev.
  Mod. Phys.}\ }\textbf {\bibinfo {volume} {32}},\ \bibinfo {pages} {335}
  (\bibinfo {year} {1960})}\BibitemShut {NoStop}%
\bibitem [{\citenamefont {Mermin}(1963)}]{NMermin63}%
  \BibitemOpen
  \bibfield  {author} {\bibinfo {author} {\bibfnamefont {N.~D.}\ \bibnamefont
  {Mermin}},\ }\href@noop {} {\bibfield  {journal} {\bibinfo  {journal} {Annals
  of Physics}\ }\textbf {\bibinfo {volume} {21}},\ \bibinfo {pages} {99}
  (\bibinfo {year} {1963})}\BibitemShut {NoStop}%
\bibitem [{\citenamefont {Weinert}\ and\ \citenamefont
  {Davenport}(1992)}]{MWeinert92}%
  \BibitemOpen
  \bibfield  {author} {\bibinfo {author} {\bibfnamefont {M.}~\bibnamefont
  {Weinert}}\ and\ \bibinfo {author} {\bibfnamefont {J.~W.}\ \bibnamefont
  {Davenport}},\ }\href@noop {} {\bibfield  {journal} {\bibinfo  {journal}
  {Phys. Rev. B}\ }\textbf {\bibinfo {volume} {45}},\ \bibinfo {pages} {R13709}
  (\bibinfo {year} {1992})}\BibitemShut {NoStop}%
\bibitem [{\citenamefont {Wentzcovitch}\ \emph {et~al.}(1992)\citenamefont
  {Wentzcovitch}, \citenamefont {Martins},\ and\ \citenamefont
  {Allen}}]{RWentzcovitch92}%
  \BibitemOpen
  \bibfield  {author} {\bibinfo {author} {\bibfnamefont {R.~M.}\ \bibnamefont
  {Wentzcovitch}}, \bibinfo {author} {\bibfnamefont {J.~L.}\ \bibnamefont
  {Martins}}, \ and\ \bibinfo {author} {\bibfnamefont {P.~B.}\ \bibnamefont
  {Allen}},\ }\href@noop {} {\bibfield  {journal} {\bibinfo  {journal} {Phys.
  Rev. B}\ }\textbf {\bibinfo {volume} {45}},\ \bibinfo {pages} {R11372}
  (\bibinfo {year} {1992})}\BibitemShut {NoStop}%
\bibitem [{\citenamefont {Niklasson}(2008{\natexlab{b}})}]{ANiklasson08b}%
  \BibitemOpen
  \bibfield  {author} {\bibinfo {author} {\bibfnamefont {A.~M.~N.}\
  \bibnamefont {Niklasson}},\ }\href@noop {} {\bibfield  {journal} {\bibinfo
  {journal} {J. Chem. Phys.}\ }\textbf {\bibinfo {volume} {129}},\ \bibinfo
  {pages} {244107} (\bibinfo {year} {2008}{\natexlab{b}})}\BibitemShut
  {NoStop}%
\bibitem [{\citenamefont {Pulay}\ and\ \citenamefont
  {Fogarasi}(2004)}]{PPulay04}%
  \BibitemOpen
  \bibfield  {author} {\bibinfo {author} {\bibfnamefont {P.}~\bibnamefont
  {Pulay}}\ and\ \bibinfo {author} {\bibfnamefont {G.}~\bibnamefont
  {Fogarasi}},\ }\href@noop {} {\bibfield  {journal} {\bibinfo  {journal}
  {Chem. Phys. Lett.}\ }\textbf {\bibinfo {volume} {386}},\ \bibinfo {pages}
  {272} (\bibinfo {year} {2004})}\BibitemShut {NoStop}%
\bibitem [{\citenamefont {Niklasson}\ \emph {et~al.}(2007)\citenamefont
  {Niklasson}, \citenamefont {Tymczak},\ and\ \citenamefont
  {Challacombe}}]{ANiklasson07}%
  \BibitemOpen
  \bibfield  {author} {\bibinfo {author} {\bibfnamefont {A.~M.~N.}\
  \bibnamefont {Niklasson}}, \bibinfo {author} {\bibfnamefont {C.~J.}\
  \bibnamefont {Tymczak}}, \ and\ \bibinfo {author} {\bibfnamefont
  {M.}~\bibnamefont {Challacombe}},\ }\href@noop {} {\bibfield  {journal}
  {\bibinfo  {journal} {J. Chem. Phys.}\ }\textbf {\bibinfo {volume} {126}},\
  \bibinfo {pages} {144103} (\bibinfo {year} {2007})}\BibitemShut {NoStop}%
\bibitem [{\citenamefont {Goedecker}\ and\ \citenamefont
  {Colombo}(1994)}]{SGoedecker94}%
  \BibitemOpen
  \bibfield  {author} {\bibinfo {author} {\bibfnamefont {S.}~\bibnamefont
  {Goedecker}}\ and\ \bibinfo {author} {\bibfnamefont {L.}~\bibnamefont
  {Colombo}},\ }\href@noop {} {\bibfield  {journal} {\bibinfo  {journal} {Phys.
  Rev. Lett.}\ }\textbf {\bibinfo {volume} {73}},\ \bibinfo {pages} {122}
  (\bibinfo {year} {1994})}\BibitemShut {NoStop}%
\bibitem [{\citenamefont {Silver}\ and\ \citenamefont
  {Roder}(1994)}]{RSilver94}%
  \BibitemOpen
  \bibfield  {author} {\bibinfo {author} {\bibfnamefont {R.~N.}\ \bibnamefont
  {Silver}}\ and\ \bibinfo {author} {\bibfnamefont {H.}~\bibnamefont {Roder}},\
  }\href@noop {} {\bibfield  {journal} {\bibinfo  {journal} {Int.\ J.\ Mod.\
  Phys.\ C}\ }\textbf {\bibinfo {volume} {5}},\ \bibinfo {pages} {735}
  (\bibinfo {year} {1994})}\BibitemShut {NoStop}%
\bibitem [{\citenamefont {Silver}\ \emph {et~al.}(1996)\citenamefont {Silver},
  \citenamefont {Roder}, \citenamefont {Voter},\ and\ \citenamefont
  {Kress}}]{RSilver96}%
  \BibitemOpen
  \bibfield  {author} {\bibinfo {author} {\bibfnamefont {R.~N.}\ \bibnamefont
  {Silver}}, \bibinfo {author} {\bibfnamefont {H.}~\bibnamefont {Roder}},
  \bibinfo {author} {\bibfnamefont {A.~F.}\ \bibnamefont {Voter}}, \ and\
  \bibinfo {author} {\bibfnamefont {J.~D.}\ \bibnamefont {Kress}},\ }\href@noop
  {} {\bibfield  {journal} {\bibinfo  {journal} {Int.\ J.\ Comput.\ Phys.}\
  }\textbf {\bibinfo {volume} {124}},\ \bibinfo {pages} {115} (\bibinfo {year}
  {1996})}\BibitemShut {NoStop}%
\bibitem [{\citenamefont {Weisse}\ \emph {et~al.}(2006)\citenamefont {Weisse},
  \citenamefont {Wellein}, \citenamefont {Alvermann},\ and\ \citenamefont
  {Fehske}}]{AWeisse06}%
  \BibitemOpen
  \bibfield  {author} {\bibinfo {author} {\bibfnamefont {A.}~\bibnamefont
  {Weisse}}, \bibinfo {author} {\bibfnamefont {G.}~\bibnamefont {Wellein}},
  \bibinfo {author} {\bibfnamefont {A.}~\bibnamefont {Alvermann}}, \ and\
  \bibinfo {author} {\bibfnamefont {H.}~\bibnamefont {Fehske}},\ }\href@noop {}
  {\bibfield  {journal} {\bibinfo  {journal} {Rev. Mod. Phys.}\ }\textbf
  {\bibinfo {volume} {78}},\ \bibinfo {pages} {275} (\bibinfo {year}
  {2006})}\BibitemShut {NoStop}%
\bibitem [{\citenamefont {Niklasson}(2003)}]{ANiklasson03B}%
  \BibitemOpen
  \bibfield  {author} {\bibinfo {author} {\bibfnamefont {A.~M.~N.}\
  \bibnamefont {Niklasson}},\ }\href@noop {} {\bibfield  {journal} {\bibinfo
  {journal} {Phys. Rev. B}\ }\textbf {\bibinfo {volume} {68}},\ \bibinfo
  {pages} {233104} (\bibinfo {year} {2003})}\BibitemShut {NoStop}%
\bibitem [{\citenamefont {Niklasson}\ \emph {et~al.}(2015)\citenamefont
  {Niklasson}, \citenamefont {Cawkwell}, \citenamefont {Rubensson},\ and\
  \citenamefont {Rudberg}}]{ANiklasson15}%
  \BibitemOpen
  \bibfield  {author} {\bibinfo {author} {\bibfnamefont {A.~M.~N.}\
  \bibnamefont {Niklasson}}, \bibinfo {author} {\bibfnamefont {M.~J.}\
  \bibnamefont {Cawkwell}}, \bibinfo {author} {\bibfnamefont {E.~H.}\
  \bibnamefont {Rubensson}}, \ and\ \bibinfo {author} {\bibfnamefont
  {E.}~\bibnamefont {Rudberg}},\ }\href@noop {} {\bibfield  {journal} {\bibinfo
   {journal} {Phys. Rev. E}\ }\textbf {\bibinfo {volume} {92}},\ \bibinfo
  {pages} {063301} (\bibinfo {year} {2015})}\BibitemShut {NoStop}%
\bibitem [{\citenamefont {Goedecker}(1999)}]{SGoedecker99}%
  \BibitemOpen
  \bibfield  {author} {\bibinfo {author} {\bibfnamefont {S.}~\bibnamefont
  {Goedecker}},\ }\href@noop {} {\bibfield  {journal} {\bibinfo  {journal}
  {Rev. Mod. Phys.}\ }\textbf {\bibinfo {volume} {71}},\ \bibinfo {pages}
  {1085} (\bibinfo {year} {1999})}\BibitemShut {NoStop}%
\bibitem [{\citenamefont {Bowler}\ and\ \citenamefont
  {Miyazaki}(2012)}]{DBowler12}%
  \BibitemOpen
  \bibfield  {author} {\bibinfo {author} {\bibfnamefont {D.~R.}\ \bibnamefont
  {Bowler}}\ and\ \bibinfo {author} {\bibfnamefont {T.}~\bibnamefont
  {Miyazaki}},\ }\href@noop {} {\bibfield  {journal} {\bibinfo  {journal} {Rep.
  Prog. Phys.}\ }\textbf {\bibinfo {volume} {75}},\ \bibinfo {pages} {036503}
  (\bibinfo {year} {2012})}\BibitemShut {NoStop}%
\bibitem [{\citenamefont {Heitler}\ and\ \citenamefont
  {London}(1927)}]{WHeitler27}%
  \BibitemOpen
  \bibfield  {author} {\bibinfo {author} {\bibfnamefont {W.}~\bibnamefont
  {Heitler}}\ and\ \bibinfo {author} {\bibfnamefont {F.}~\bibnamefont
  {London}},\ }\href@noop {} {\bibfield  {journal} {\bibinfo  {journal} {Z.\
  Phys.}\ }\textbf {\bibinfo {volume} {44}},\ \bibinfo {pages} {455} (\bibinfo
  {year} {1927})}\BibitemShut {NoStop}%
\bibitem [{\citenamefont {Born}\ and\ \citenamefont
  {Oppenheimer}(1927)}]{MBorn27}%
  \BibitemOpen
  \bibfield  {author} {\bibinfo {author} {\bibfnamefont {M.}~\bibnamefont
  {Born}}\ and\ \bibinfo {author} {\bibfnamefont {R.}~\bibnamefont
  {Oppenheimer}},\ }\href@noop {} {\bibfield  {journal} {\bibinfo  {journal}
  {Ann.\ Phys.}\ }\textbf {\bibinfo {volume} {389}},\ \bibinfo {pages} {475}
  (\bibinfo {year} {1927})}\BibitemShut {NoStop}%
\bibitem [{\citenamefont {Odell}\ \emph {et~al.}(2009)\citenamefont {Odell},
  \citenamefont {Delin}, \citenamefont {Johansson}, \citenamefont {Bock},
  \citenamefont {Challacombe},\ and\ \citenamefont {Niklasson}}]{AOdell09}%
  \BibitemOpen
  \bibfield  {author} {\bibinfo {author} {\bibfnamefont {A.}~\bibnamefont
  {Odell}}, \bibinfo {author} {\bibfnamefont {A.}~\bibnamefont {Delin}},
  \bibinfo {author} {\bibfnamefont {B.}~\bibnamefont {Johansson}}, \bibinfo
  {author} {\bibfnamefont {N.}~\bibnamefont {Bock}}, \bibinfo {author}
  {\bibfnamefont {M.}~\bibnamefont {Challacombe}}, \ and\ \bibinfo {author}
  {\bibfnamefont {A.~M.~N.}\ \bibnamefont {Niklasson}},\ }\href@noop {}
  {\bibfield  {journal} {\bibinfo  {journal} {J. Chem. Phys.}\ }\textbf
  {\bibinfo {volume} {131}},\ \bibinfo {pages} {244106} (\bibinfo {year}
  {2009})}\BibitemShut {NoStop}%
\bibitem [{\citenamefont {Odell}\ \emph {et~al.}(2011)\citenamefont {Odell},
  \citenamefont {Delin}, \citenamefont {Johansson}, \citenamefont {Cawkwell},\
  and\ \citenamefont {Niklasson}}]{AOdell11}%
  \BibitemOpen
  \bibfield  {author} {\bibinfo {author} {\bibfnamefont {A.}~\bibnamefont
  {Odell}}, \bibinfo {author} {\bibfnamefont {A.}~\bibnamefont {Delin}},
  \bibinfo {author} {\bibfnamefont {B.}~\bibnamefont {Johansson}}, \bibinfo
  {author} {\bibfnamefont {M.~J.}\ \bibnamefont {Cawkwell}}, \ and\ \bibinfo
  {author} {\bibfnamefont {A.~M.~N.}\ \bibnamefont {Niklasson}},\ }\href@noop
  {} {\bibfield  {journal} {\bibinfo  {journal} {J. Chem. Phys.}\ }\textbf
  {\bibinfo {volume} {135}},\ \bibinfo {pages} {224105} (\bibinfo {year}
  {2011})}\BibitemShut {NoStop}%
\bibitem [{\citenamefont {Knoll}\ and\ \citenamefont
  {Keyes}(2004)}]{DAKnoll04}%
  \BibitemOpen
  \bibfield  {author} {\bibinfo {author} {\bibfnamefont {D.}~\bibnamefont
  {Knoll}}\ and\ \bibinfo {author} {\bibfnamefont {D.}~\bibnamefont {Keyes}},\
  }\href@noop {} {\bibfield  {journal} {\bibinfo  {journal} {J. Comput. Phys.}\
  }\textbf {\bibinfo {volume} {193}},\ \bibinfo {pages} {357} (\bibinfo {year}
  {2004})}\BibitemShut {NoStop}%
\bibitem [{\citenamefont {Anderson}(1965)}]{DGAnderson65}%
  \BibitemOpen
  \bibfield  {author} {\bibinfo {author} {\bibfnamefont {D.~G.}\ \bibnamefont
  {Anderson}},\ }\href@noop {} {\bibfield  {journal} {\bibinfo  {journal} {J.
  Assoc. Comput. Mach.}\ }\textbf {\bibinfo {volume} {12}},\ \bibinfo {pages}
  {547} (\bibinfo {year} {1965})}\BibitemShut {NoStop}%
\bibitem [{\citenamefont {Saad}\ and\ \citenamefont {Schultz}(1986)}]{YSaad86}%
  \BibitemOpen
  \bibfield  {author} {\bibinfo {author} {\bibfnamefont {Y.}~\bibnamefont
  {Saad}}\ and\ \bibinfo {author} {\bibfnamefont {M.~H.}\ \bibnamefont
  {Schultz}},\ }\href@noop {} {\bibfield  {journal} {\bibinfo  {journal} {SIAM
  J. Sci. Stat. Comput.}\ }\textbf {\bibinfo {volume} {7}},\ \bibinfo {pages}
  {856} (\bibinfo {year} {1986})}\BibitemShut {NoStop}%
\bibitem [{\citenamefont {Saad}(1996)}]{YSaad96}%
  \BibitemOpen
  \bibfield  {author} {\bibinfo {author} {\bibfnamefont {Y.}~\bibnamefont
  {Saad}},\ }\href@noop {} {\emph {\bibinfo {title} {Iterative methods for
  Sparse Linear Systems}}}\ (\bibinfo  {publisher} {PWS Publishing},\ \bibinfo
  {address} {Boston},\ \bibinfo {year} {1996})\BibitemShut {NoStop}%
\bibitem [{\citenamefont {Niklasson}(2020{\natexlab{b}})}]{NiklassonUnPub}%
  \BibitemOpen
  \bibfield  {author} {\bibinfo {author} {\bibfnamefont {A.~M.~N.}\
  \bibnamefont {Niklasson}},\ }\href@noop {} {\enquote {\bibinfo {title}
  {Preconditioning in extended lagrangian born-oppenheimer molecular
  dynamics},}\ } (\bibinfo {year} {2020}{\natexlab{b}}),\ \bibinfo {note}
  {unpublished}\BibitemShut {NoStop}%
\bibitem [{\citenamefont {Niklasson}\ and\ \citenamefont
  {Challacombe}(2004)}]{ANiklasson04}%
  \BibitemOpen
  \bibfield  {author} {\bibinfo {author} {\bibfnamefont {A.~M.~N.}\
  \bibnamefont {Niklasson}}\ and\ \bibinfo {author} {\bibfnamefont
  {M.}~\bibnamefont {Challacombe}},\ }\href@noop {} {\bibfield  {journal}
  {\bibinfo  {journal} {Phys. Rev. Lett.}\ }\textbf {\bibinfo {volume} {92}},\
  \bibinfo {pages} {193001} (\bibinfo {year} {2004})}\BibitemShut {NoStop}%
\bibitem [{\citenamefont {Adler}(1962)}]{SLAdler62}%
  \BibitemOpen
  \bibfield  {author} {\bibinfo {author} {\bibfnamefont {S.~L.}\ \bibnamefont
  {Adler}},\ }\href@noop {} {\bibfield  {journal} {\bibinfo  {journal} {Phys.
  Rev.}\ }\textbf {\bibinfo {volume} {126}},\ \bibinfo {pages} {413} (\bibinfo
  {year} {1962})}\BibitemShut {NoStop}%
\bibitem [{\citenamefont {Wiser}(1963)}]{NWiser63}%
  \BibitemOpen
  \bibfield  {author} {\bibinfo {author} {\bibfnamefont {N.}~\bibnamefont
  {Wiser}},\ }\href@noop {} {\bibfield  {journal} {\bibinfo  {journal} {Phys.
  Rev.}\ }\textbf {\bibinfo {volume} {129}},\ \bibinfo {pages} {62} (\bibinfo
  {year} {1963})}\BibitemShut {NoStop}%
\bibitem [{\citenamefont {Nishimoto}(2017)}]{YNishimoto17}%
  \BibitemOpen
  \bibfield  {author} {\bibinfo {author} {\bibfnamefont {Y.}~\bibnamefont
  {Nishimoto}},\ }\href@noop {} {\bibfield  {journal} {\bibinfo  {journal} {J.
  Chem. Phys.}\ }\textbf {\bibinfo {volume} {146}},\ \bibinfo {pages} {084101}
  (\bibinfo {year} {2017})}\BibitemShut {NoStop}%
\bibitem [{\citenamefont {Pulay}(1969)}]{PPulay69}%
  \BibitemOpen
  \bibfield  {author} {\bibinfo {author} {\bibfnamefont {P.}~\bibnamefont
  {Pulay}},\ }\href@noop {} {\bibfield  {journal} {\bibinfo  {journal} {Mol.
  Phys.}\ }\textbf {\bibinfo {volume} {17}},\ \bibinfo {pages} {197} (\bibinfo
  {year} {1969})}\BibitemShut {NoStop}%
\bibitem [{\citenamefont {Niklasson}(2004)}]{ANiklasson04B}%
  \BibitemOpen
  \bibfield  {author} {\bibinfo {author} {\bibfnamefont {A.~M.~N.}\
  \bibnamefont {Niklasson}},\ }\href@noop {} {\bibfield  {journal} {\bibinfo
  {journal} {Phys. Rev. B}\ }\textbf {\bibinfo {volume} {70}},\ \bibinfo
  {pages} {193102} (\bibinfo {year} {2004})}\BibitemShut {NoStop}%
\bibitem [{\citenamefont {Thijssen}(1999)}]{JMThijssen99}%
  \BibitemOpen
  \bibfield  {author} {\bibinfo {author} {\bibfnamefont {J.~M.}\ \bibnamefont
  {Thijssen}},\ }\enquote {\bibinfo {title} {Computational physics},}\ in\
  \href@noop {} {\emph {\bibinfo {booktitle} {Computational Physics}}}\
  (\bibinfo  {publisher} {Cambride University Press},\ \bibinfo {year}
  {1999})\BibitemShut {NoStop}%
\bibitem [{\citenamefont {Elstner}\ \emph {et~al.}(1998)\citenamefont
  {Elstner}, \citenamefont {Poresag}, \citenamefont {Jungnickel}, \citenamefont
  {Elsner}, \citenamefont {Haugk}, \citenamefont {Frauenheim}, \citenamefont
  {Suhai},\ and\ \citenamefont {Seifert}}]{MElstner98}%
  \BibitemOpen
  \bibfield  {author} {\bibinfo {author} {\bibfnamefont {M.}~\bibnamefont
  {Elstner}}, \bibinfo {author} {\bibfnamefont {D.}~\bibnamefont {Poresag}},
  \bibinfo {author} {\bibfnamefont {G.}~\bibnamefont {Jungnickel}}, \bibinfo
  {author} {\bibfnamefont {J.}~\bibnamefont {Elsner}}, \bibinfo {author}
  {\bibfnamefont {M.}~\bibnamefont {Haugk}}, \bibinfo {author} {\bibfnamefont
  {T.}~\bibnamefont {Frauenheim}}, \bibinfo {author} {\bibfnamefont
  {S.}~\bibnamefont {Suhai}}, \ and\ \bibinfo {author} {\bibfnamefont
  {G.}~\bibnamefont {Seifert}},\ }\href@noop {} {\bibfield  {journal} {\bibinfo
   {journal} {Phys. Rev. B}\ }\textbf {\bibinfo {volume} {58}},\ \bibinfo
  {pages} {7260} (\bibinfo {year} {1998})}\BibitemShut {NoStop}%
\bibitem [{\citenamefont {Finnis}\ \emph {et~al.}(1998)\citenamefont {Finnis},
  \citenamefont {Paxton}, \citenamefont {Methfessel},\ and\ \citenamefont {van
  Schilfgarde}}]{MFinnis98}%
  \BibitemOpen
  \bibfield  {author} {\bibinfo {author} {\bibfnamefont {M.~W.}\ \bibnamefont
  {Finnis}}, \bibinfo {author} {\bibfnamefont {A.~T.}\ \bibnamefont {Paxton}},
  \bibinfo {author} {\bibfnamefont {M.}~\bibnamefont {Methfessel}}, \ and\
  \bibinfo {author} {\bibfnamefont {M.}~\bibnamefont {van Schilfgarde}},\
  }\href@noop {} {\bibfield  {journal} {\bibinfo  {journal} {Phys. Rev. Lett.}\
  }\textbf {\bibinfo {volume} {81}},\ \bibinfo {pages} {5149} (\bibinfo {year}
  {1998})}\BibitemShut {NoStop}%
\bibitem [{\citenamefont {Frauenheim}\ \emph {et~al.}(2000)\citenamefont
  {Frauenheim}, \citenamefont {Seifert}, \citenamefont {aand Z.~Hajnal},
  \citenamefont {Jungnickel}, \citenamefont {Poresag}, \citenamefont {Suhai},\
  and\ \citenamefont {Scholz}}]{TFrauenheim00}%
  \BibitemOpen
  \bibfield  {author} {\bibinfo {author} {\bibfnamefont {T.}~\bibnamefont
  {Frauenheim}}, \bibinfo {author} {\bibfnamefont {G.}~\bibnamefont {Seifert}},
  \bibinfo {author} {\bibfnamefont {M.~E.}\ \bibnamefont {aand Z.~Hajnal}},
  \bibinfo {author} {\bibfnamefont {G.}~\bibnamefont {Jungnickel}}, \bibinfo
  {author} {\bibfnamefont {D.}~\bibnamefont {Poresag}}, \bibinfo {author}
  {\bibfnamefont {S.}~\bibnamefont {Suhai}}, \ and\ \bibinfo {author}
  {\bibfnamefont {R.}~\bibnamefont {Scholz}},\ }\href@noop {} {\bibfield
  {journal} {\bibinfo  {journal} {Phys. Stat. sol.}\ }\textbf {\bibinfo
  {volume} {217}},\ \bibinfo {pages} {41} (\bibinfo {year} {2000})}\BibitemShut
  {NoStop}%
\bibitem [{\citenamefont {Gaus}\ \emph {et~al.}(2011)\citenamefont {Gaus},
  \citenamefont {Cui},\ and\ \citenamefont {Elstner}}]{MGaus11}%
  \BibitemOpen
  \bibfield  {author} {\bibinfo {author} {\bibfnamefont {M.}~\bibnamefont
  {Gaus}}, \bibinfo {author} {\bibfnamefont {Q.}~\bibnamefont {Cui}}, \ and\
  \bibinfo {author} {\bibfnamefont {M.}~\bibnamefont {Elstner}},\ }\href@noop
  {} {\bibfield  {journal} {\bibinfo  {journal} {J, Chem. Theory Comput.}\
  }\textbf {\bibinfo {volume} {7}},\ \bibinfo {pages} {931} (\bibinfo {year}
  {2011})}\BibitemShut {NoStop}%
\bibitem [{\citenamefont {et~al.}(2020)}]{BHourahine20}%
  \BibitemOpen
  \bibfield  {author} {\bibinfo {author} {\bibfnamefont {B.~H.}\ \bibnamefont
  {et~al.}},\ }\href@noop {} {\bibfield  {journal} {\bibinfo  {journal} {J.
  Chem. Phys.}\ }\textbf {\bibinfo {volume} {152}},\ \bibinfo {pages} {0000}
  (\bibinfo {year} {2020})}\BibitemShut {NoStop}%
\bibitem [{\citenamefont {Cawkwell}\ and\ \citenamefont
  {et~al.}(2010)}]{LATTE}%
  \BibitemOpen
  \bibfield  {author} {\bibinfo {author} {\bibfnamefont {M.~J.}\ \bibnamefont
  {Cawkwell}}\ and\ \bibinfo {author} {\bibnamefont {et~al.}},\ }\href@noop {}
  {\enquote {\bibinfo {title} {{\sc LATTE}},}\ } (\bibinfo {year} {2010}),\
  \bibinfo {note} {\mbox{L}os Alamos National Laboratory (LA- CC-10004),
  http://www.github.com/lanl/latte}\BibitemShut {NoStop}%
\bibitem [{\citenamefont {Krishnapryian}\ \emph {et~al.}(2017)\citenamefont
  {Krishnapryian}, \citenamefont {Yang}, \citenamefont {Niklasson},\ and\
  \citenamefont {Cawkwell}}]{AKrishnapriyan17}%
  \BibitemOpen
  \bibfield  {author} {\bibinfo {author} {\bibfnamefont {A.}~\bibnamefont
  {Krishnapryian}}, \bibinfo {author} {\bibfnamefont {P.}~\bibnamefont {Yang}},
  \bibinfo {author} {\bibfnamefont {A.~M.~N.}\ \bibnamefont {Niklasson}}, \
  and\ \bibinfo {author} {\bibfnamefont {M.~J.}\ \bibnamefont {Cawkwell}},\
  }\href@noop {} {\bibfield  {journal} {\bibinfo  {journal} {J. Chem. Theory
  Comput.}\ }\textbf {\bibinfo {volume} {13}},\ \bibinfo {pages} {6191}
  (\bibinfo {year} {2017})}\BibitemShut {NoStop}%
\bibitem [{\citenamefont {Schmidt}\ \emph {et~al.}(1993)\citenamefont
  {Schmidt}, \citenamefont {Baldridge}, \citenamefont {Boatz}, \citenamefont
  {Elbert}, \citenamefont {Gordon}, \citenamefont {Jensen}, \citenamefont
  {Koseki}, \citenamefont {Matsunaga}, \citenamefont {Nguyen}, \citenamefont
  {Su}, \citenamefont {Windus}, \citenamefont {Dupuis},\ and\ \citenamefont
  {Montgomery}}]{gamess}%
  \BibitemOpen
  \bibfield  {author} {\bibinfo {author} {\bibfnamefont {M.~W.}\ \bibnamefont
  {Schmidt}}, \bibinfo {author} {\bibfnamefont {K.~K.}\ \bibnamefont
  {Baldridge}}, \bibinfo {author} {\bibfnamefont {J.~A.}\ \bibnamefont
  {Boatz}}, \bibinfo {author} {\bibfnamefont {S.~T.}\ \bibnamefont {Elbert}},
  \bibinfo {author} {\bibfnamefont {M.~S.}\ \bibnamefont {Gordon}}, \bibinfo
  {author} {\bibfnamefont {J.~H.}\ \bibnamefont {Jensen}}, \bibinfo {author}
  {\bibfnamefont {S.}~\bibnamefont {Koseki}}, \bibinfo {author} {\bibfnamefont
  {N.}~\bibnamefont {Matsunaga}}, \bibinfo {author} {\bibfnamefont {K.~A.}\
  \bibnamefont {Nguyen}}, \bibinfo {author} {\bibfnamefont {S.}~\bibnamefont
  {Su}}, \bibinfo {author} {\bibfnamefont {T.~L.}\ \bibnamefont {Windus}},
  \bibinfo {author} {\bibfnamefont {M.}~\bibnamefont {Dupuis}}, \ and\ \bibinfo
  {author} {\bibfnamefont {J.~A.}\ \bibnamefont {Montgomery}},\ }\href@noop {}
  {\bibfield  {journal} {\bibinfo  {journal} {J. Comp. Chem.}\ }\textbf
  {\bibinfo {volume} {14}},\ \bibinfo {pages} {1347} (\bibinfo {year}
  {1993})}\BibitemShut {NoStop}%
\bibitem [{\citenamefont {Frisch}\ \emph {et~al.}(1994)\citenamefont {Frisch},
  \citenamefont {Trucks}, \citenamefont {Schlegel}, \citenamefont {Gill},
  \citenamefont {Johnson}, \citenamefont {Robb}, \citenamefont {Cheeseman},
  \citenamefont {Keith}, \citenamefont {Petersson}, \citenamefont {Montgomery},
  \citenamefont {Raghavachari}, \citenamefont {Al-Laham}, \citenamefont
  {Zakrzewski}, \citenamefont {Ortiz}, \citenamefont {Foresman}, \citenamefont
  {Peng}, \citenamefont {Ayala}, \citenamefont {Chen}, \citenamefont {Wong},
  \citenamefont {Andres}, \citenamefont {Replogle}, \citenamefont {Gomperts},
  \citenamefont {Martin}, \citenamefont {Fox}, \citenamefont {Binkley},
  \citenamefont {Defrees}, \citenamefont {Baker}, \citenamefont {Stewart},
  \citenamefont {Head-Gordon}, \citenamefont {Gonzalez},\ and\ \citenamefont
  {Pople}}]{Gaussian94A}%
  \BibitemOpen
  \bibfield  {author} {\bibinfo {author} {\bibfnamefont {M.~J.}\ \bibnamefont
  {Frisch}}, \bibinfo {author} {\bibfnamefont {G.~W.}\ \bibnamefont {Trucks}},
  \bibinfo {author} {\bibfnamefont {H.~B.}\ \bibnamefont {Schlegel}}, \bibinfo
  {author} {\bibfnamefont {P.~M.~W.}\ \bibnamefont {Gill}}, \bibinfo {author}
  {\bibfnamefont {B.~G.}\ \bibnamefont {Johnson}}, \bibinfo {author}
  {\bibfnamefont {M.~A.}\ \bibnamefont {Robb}}, \bibinfo {author}
  {\bibfnamefont {J.~R.}\ \bibnamefont {Cheeseman}}, \bibinfo {author}
  {\bibfnamefont {T.}~\bibnamefont {Keith}}, \bibinfo {author} {\bibfnamefont
  {G.~A.}\ \bibnamefont {Petersson}}, \bibinfo {author} {\bibfnamefont {J.~A.}\
  \bibnamefont {Montgomery}}, \bibinfo {author} {\bibfnamefont
  {K.}~\bibnamefont {Raghavachari}}, \bibinfo {author} {\bibfnamefont {M.~A.}\
  \bibnamefont {Al-Laham}}, \bibinfo {author} {\bibfnamefont {V.~G.}\
  \bibnamefont {Zakrzewski}}, \bibinfo {author} {\bibfnamefont {J.~V.}\
  \bibnamefont {Ortiz}}, \bibinfo {author} {\bibfnamefont {J.~B.}\ \bibnamefont
  {Foresman}}, \bibinfo {author} {\bibfnamefont {C.~Y.}\ \bibnamefont {Peng}},
  \bibinfo {author} {\bibfnamefont {P.~Y.}\ \bibnamefont {Ayala}}, \bibinfo
  {author} {\bibfnamefont {W.}~\bibnamefont {Chen}}, \bibinfo {author}
  {\bibfnamefont {M.~W.}\ \bibnamefont {Wong}}, \bibinfo {author}
  {\bibfnamefont {J.~L.}\ \bibnamefont {Andres}}, \bibinfo {author}
  {\bibfnamefont {E.~S.}\ \bibnamefont {Replogle}}, \bibinfo {author}
  {\bibfnamefont {R.}~\bibnamefont {Gomperts}}, \bibinfo {author}
  {\bibfnamefont {R.~L.}\ \bibnamefont {Martin}}, \bibinfo {author}
  {\bibfnamefont {D.~J.}\ \bibnamefont {Fox}}, \bibinfo {author} {\bibfnamefont
  {J.~S.}\ \bibnamefont {Binkley}}, \bibinfo {author} {\bibfnamefont {D.~J.}\
  \bibnamefont {Defrees}}, \bibinfo {author} {\bibfnamefont {J.}~\bibnamefont
  {Baker}}, \bibinfo {author} {\bibfnamefont {J.~P.}\ \bibnamefont {Stewart}},
  \bibinfo {author} {\bibfnamefont {M.}~\bibnamefont {Head-Gordon}}, \bibinfo
  {author} {\bibfnamefont {C.}~\bibnamefont {Gonzalez}}, \ and\ \bibinfo
  {author} {\bibfnamefont {J.~A.}\ \bibnamefont {Pople}},\ }\href@noop {}
  {\emph {\bibinfo {title} {Gaussian 94, \mbox{Revision B.3}}}},\ \bibinfo
  {organization} {{\sc Gaussian}} (\bibinfo {year} {1994})\BibitemShut
  {NoStop}%
\bibitem [{\citenamefont {Stewart}(1996)}]{JStewart96}%
  \BibitemOpen
  \bibfield  {author} {\bibinfo {author} {\bibfnamefont {J.~P.}\ \bibnamefont
  {Stewart}},\ }\href@noop {} {\bibfield  {journal} {\bibinfo  {journal} {Int.
  J. Quant. Chem.}\ }\textbf {\bibinfo {volume} {58}},\ \bibinfo {pages} {133}
  (\bibinfo {year} {1996})}\BibitemShut {NoStop}%
\bibitem [{\citenamefont {Soler}\ \emph {et~al.}(2002)\citenamefont {Soler},
  \citenamefont {Artacho}, \citenamefont {Gale}, \citenamefont {Garcia},
  \citenamefont {Junquera}, \citenamefont {Ordejon},\ and\ \citenamefont
  {Sanchez-Portal}}]{JSoler02}%
  \BibitemOpen
  \bibfield  {author} {\bibinfo {author} {\bibfnamefont {J.~M.}\ \bibnamefont
  {Soler}}, \bibinfo {author} {\bibfnamefont {E.}~\bibnamefont {Artacho}},
  \bibinfo {author} {\bibfnamefont {J.~D.}\ \bibnamefont {Gale}}, \bibinfo
  {author} {\bibfnamefont {A.}~\bibnamefont {Garcia}}, \bibinfo {author}
  {\bibfnamefont {J.}~\bibnamefont {Junquera}}, \bibinfo {author}
  {\bibfnamefont {P.}~\bibnamefont {Ordejon}}, \ and\ \bibinfo {author}
  {\bibfnamefont {D.}~\bibnamefont {Sanchez-Portal}},\ }\href@noop {}
  {\bibfield  {journal} {\bibinfo  {journal} {J. Phys.: Condens. Matter}\
  }\textbf {\bibinfo {volume} {14}},\ \bibinfo {pages} {2745} (\bibinfo {year}
  {2002})}\BibitemShut {NoStop}%
\bibitem [{\citenamefont {Bowler}\ \emph {et~al.}(2006)\citenamefont {Bowler},
  \citenamefont {Choudhury}, \citenamefont {Gillan},\ and\ \citenamefont
  {Miyazaki}}]{DBowler06}%
  \BibitemOpen
  \bibfield  {author} {\bibinfo {author} {\bibfnamefont {D.~R.}\ \bibnamefont
  {Bowler}}, \bibinfo {author} {\bibfnamefont {R.}~\bibnamefont {Choudhury}},
  \bibinfo {author} {\bibfnamefont {M.~J.}\ \bibnamefont {Gillan}}, \ and\
  \bibinfo {author} {\bibfnamefont {T.}~\bibnamefont {Miyazaki}},\ }\href@noop
  {} {\bibfield  {journal} {\bibinfo  {journal} {Phys. Stat. Sol. B}\ }\textbf
  {\bibinfo {volume} {243}},\ \bibinfo {pages} {898} (\bibinfo {year}
  {2006})}\BibitemShut {NoStop}%
\bibitem [{\citenamefont {Hine}\ \emph {et~al.}(2009)\citenamefont {Hine},
  \citenamefont {Haynes}, \citenamefont {Mostofi}, \citenamefont {Skylaris},\
  and\ \citenamefont {Payne}}]{NHine09}%
  \BibitemOpen
  \bibfield  {author} {\bibinfo {author} {\bibfnamefont {N.~D.}\ \bibnamefont
  {Hine}}, \bibinfo {author} {\bibfnamefont {P.~D.}\ \bibnamefont {Haynes}},
  \bibinfo {author} {\bibfnamefont {A.~A.}\ \bibnamefont {Mostofi}}, \bibinfo
  {author} {\bibfnamefont {C.-K.}\ \bibnamefont {Skylaris}}, \ and\ \bibinfo
  {author} {\bibfnamefont {M.~C.}\ \bibnamefont {Payne}},\ }\href@noop {}
  {\bibfield  {journal} {\bibinfo  {journal} {Comput. Phys. Comm.}\ }\textbf
  {\bibinfo {volume} {180}},\ \bibinfo {pages} {1041} (\bibinfo {year}
  {2009})}\BibitemShut {NoStop}%
\bibitem [{\citenamefont {Kendall}\ \emph {et~al.}(2000)\citenamefont
  {Kendall}, \citenamefont {Apra}, \citenamefont {Bernholdt}, \citenamefont
  {Bylaska}, \citenamefont {Dupuis}, \citenamefont {Fann}, \citenamefont
  {Harrison}, \citenamefont {Ju}, \citenamefont {Nichols}, \citenamefont
  {Nieplocha}, \citenamefont {Straatsma}, \citenamefont {Windus},\ and\
  \citenamefont {Wong}}]{RKendall00}%
  \BibitemOpen
  \bibfield  {author} {\bibinfo {author} {\bibfnamefont {R.~A.}\ \bibnamefont
  {Kendall}}, \bibinfo {author} {\bibfnamefont {E.}~\bibnamefont {Apra}},
  \bibinfo {author} {\bibfnamefont {D.~E.}\ \bibnamefont {Bernholdt}}, \bibinfo
  {author} {\bibfnamefont {E.~J.}\ \bibnamefont {Bylaska}}, \bibinfo {author}
  {\bibfnamefont {M.}~\bibnamefont {Dupuis}}, \bibinfo {author} {\bibfnamefont
  {G.~I.}\ \bibnamefont {Fann}}, \bibinfo {author} {\bibfnamefont {R.~J.}\
  \bibnamefont {Harrison}}, \bibinfo {author} {\bibfnamefont {J.}~\bibnamefont
  {Ju}}, \bibinfo {author} {\bibfnamefont {J.~A.}\ \bibnamefont {Nichols}},
  \bibinfo {author} {\bibfnamefont {J.}~\bibnamefont {Nieplocha}}, \bibinfo
  {author} {\bibfnamefont {T.~P.}\ \bibnamefont {Straatsma}}, \bibinfo {author}
  {\bibfnamefont {T.~L.}\ \bibnamefont {Windus}}, \ and\ \bibinfo {author}
  {\bibfnamefont {A.~T.}\ \bibnamefont {Wong}},\ }\href@noop {} {\bibfield
  {journal} {\bibinfo  {journal} {Comput. Phys. Commun.}\ }\textbf {\bibinfo
  {volume} {128}},\ \bibinfo {pages} {260} (\bibinfo {year}
  {2000})}\BibitemShut {NoStop}%
\bibitem [{\citenamefont {VandeVondele}\ \emph {et~al.}(2005)\citenamefont
  {VandeVondele}, \citenamefont {Krack}, \citenamefont {Mohammed},
  \citenamefont {Parrinello}, \citenamefont {Chassing},\ and\ \citenamefont
  {Hutter}}]{JVandevondele05}%
  \BibitemOpen
  \bibfield  {author} {\bibinfo {author} {\bibfnamefont {J.}~\bibnamefont
  {VandeVondele}}, \bibinfo {author} {\bibfnamefont {M.}~\bibnamefont {Krack}},
  \bibinfo {author} {\bibfnamefont {F.}~\bibnamefont {Mohammed}}, \bibinfo
  {author} {\bibfnamefont {M.}~\bibnamefont {Parrinello}}, \bibinfo {author}
  {\bibfnamefont {T.}~\bibnamefont {Chassing}}, \ and\ \bibinfo {author}
  {\bibfnamefont {J.}~\bibnamefont {Hutter}},\ }\href@noop {} {\bibfield
  {journal} {\bibinfo  {journal} {Comput. Phys. Commun.}\ }\textbf {\bibinfo
  {volume} {167}},\ \bibinfo {pages} {103} (\bibinfo {year}
  {2005})}\BibitemShut {NoStop}%
\bibitem [{\citenamefont {Aradi}\ \emph {et~al.}(2007)\citenamefont {Aradi},
  \citenamefont {Hourahine},\ and\ \citenamefont {Frauenheim}}]{BAradi07}%
  \BibitemOpen
  \bibfield  {author} {\bibinfo {author} {\bibfnamefont {B.}~\bibnamefont
  {Aradi}}, \bibinfo {author} {\bibfnamefont {B.}~\bibnamefont {Hourahine}}, \
  and\ \bibinfo {author} {\bibfnamefont {T.}~\bibnamefont {Frauenheim}},\
  }\href {\doibase 10.1021/jp070186p} {\bibfield  {journal} {\bibinfo
  {journal} {The Journal of Physical Chemistry A}\ }\textbf {\bibinfo {volume}
  {111}},\ \bibinfo {pages} {5678} (\bibinfo {year} {2007})}\BibitemShut
  {NoStop}%
\bibitem [{\citenamefont {Rudberg}\ \emph {et~al.}(2011)\citenamefont
  {Rudberg}, \citenamefont {Rubensson},\ and\ \citenamefont
  {Salek}}]{ERudberg_11}%
  \BibitemOpen
  \bibfield  {author} {\bibinfo {author} {\bibfnamefont {E.}~\bibnamefont
  {Rudberg}}, \bibinfo {author} {\bibfnamefont {E.~H.}\ \bibnamefont
  {Rubensson}}, \ and\ \bibinfo {author} {\bibfnamefont {P.}~\bibnamefont
  {Salek}},\ }\href@noop {} {\bibfield  {journal} {\bibinfo  {journal} {J.
  Chem. Theory Comput.}\ }\textbf {\bibinfo {volume} {7}},\ \bibinfo {pages}
  {340} (\bibinfo {year} {2011})}\BibitemShut {NoStop}%
\bibitem [{\citenamefont {Osei-Kuffuor}\ \emph {et~al.}(2014)\citenamefont
  {Osei-Kuffuor}, \citenamefont {Fattebert},\ and\ \citenamefont
  {Gygi}}]{DOseiKuffuor14}%
  \BibitemOpen
  \bibfield  {author} {\bibinfo {author} {\bibfnamefont {D.}~\bibnamefont
  {Osei-Kuffuor}}, \bibinfo {author} {\bibfnamefont {J.~L.}\ \bibnamefont
  {Fattebert}}, \ and\ \bibinfo {author} {\bibfnamefont {F.}~\bibnamefont
  {Gygi}},\ }\href@noop {} {\bibfield  {journal} {\bibinfo  {journal} {Phys.
  Rev. Lett.}\ }\textbf {\bibinfo {volume} {112}},\ \bibinfo {pages} {046401}
  (\bibinfo {year} {2014})}\BibitemShut {NoStop}%
\bibitem [{\citenamefont {Levchenko}\ \emph {et~al.}(2015)\citenamefont
  {Levchenko}, \citenamefont {Ren}, \citenamefont {Wieferink}, \citenamefont
  {Johanni}, \citenamefont {Rinke}, \citenamefont {Blum},\ and\ \citenamefont
  {Scheffler}}]{SVLevchenko15}%
  \BibitemOpen
  \bibfield  {author} {\bibinfo {author} {\bibfnamefont {S.~V.}\ \bibnamefont
  {Levchenko}}, \bibinfo {author} {\bibfnamefont {X.}~\bibnamefont {Ren}},
  \bibinfo {author} {\bibfnamefont {J.}~\bibnamefont {Wieferink}}, \bibinfo
  {author} {\bibfnamefont {R.}~\bibnamefont {Johanni}}, \bibinfo {author}
  {\bibfnamefont {P.}~\bibnamefont {Rinke}}, \bibinfo {author} {\bibfnamefont
  {V.}~\bibnamefont {Blum}}, \ and\ \bibinfo {author} {\bibfnamefont
  {M.}~\bibnamefont {Scheffler}},\ }\href@noop {} {\bibfield  {journal}
  {\bibinfo  {journal} {Comput. Phys. Comm.}\ }\textbf {\bibinfo {volume}
  {192}},\ \bibinfo {pages} {60} (\bibinfo {year} {2015})}\BibitemShut
  {NoStop}%
\bibitem [{\citenamefont {Martinez}\ \emph {et~al.}(2015)\citenamefont
  {Martinez}, \citenamefont {Cawkwell}, , \citenamefont {Voter},\ and\
  \citenamefont {Niklasson}}]{EMartinez15}%
  \BibitemOpen
  \bibfield  {author} {\bibinfo {author} {\bibfnamefont {E.}~\bibnamefont
  {Martinez}}, \bibinfo {author} {\bibfnamefont {M.~J.}\ \bibnamefont
  {Cawkwell}}, , \bibinfo {author} {\bibfnamefont {A.~F.}\ \bibnamefont
  {Voter}}, \ and\ \bibinfo {author} {\bibfnamefont {A.~M.~N.}\ \bibnamefont
  {Niklasson}},\ }\href@noop {} {\bibfield  {journal} {\bibinfo  {journal} {J.
  Chem. Phys.}\ }\textbf {\bibinfo {volume} {142}},\ \bibinfo {pages} {1770}
  (\bibinfo {year} {2015})}\BibitemShut {NoStop}%
\bibitem [{\citenamefont {Kerker}(1981)}]{GPKerker81}%
  \BibitemOpen
  \bibfield  {author} {\bibinfo {author} {\bibfnamefont {G.~P.}\ \bibnamefont
  {Kerker}},\ }\href {\doibase 10.1103/PhysRevB.23.3082} {\bibfield  {journal}
  {\bibinfo  {journal} {Phys. Rev. B}\ }\textbf {\bibinfo {volume} {23}},\
  \bibinfo {pages} {3082} (\bibinfo {year} {1981})}\BibitemShut {NoStop}%
\bibitem [{\citenamefont {Herbert}\ and\ \citenamefont
  {Head-Gordon}(2005)}]{JMHerbert05}%
  \BibitemOpen
  \bibfield  {author} {\bibinfo {author} {\bibfnamefont {J.}~\bibnamefont
  {Herbert}}\ and\ \bibinfo {author} {\bibfnamefont {M.}~\bibnamefont
  {Head-Gordon}},\ }\href@noop {} {\bibfield  {journal} {\bibinfo  {journal}
  {Phys. Chem. Chem. Phys.}\ }\textbf {\bibinfo {volume} {7}},\ \bibinfo
  {pages} {3269} (\bibinfo {year} {2005})}\BibitemShut {NoStop}%
\bibitem [{\citenamefont {K\"{u}hne}\ \emph {et~al.}(2007)\citenamefont
  {K\"{u}hne}, \citenamefont {Krack}, \citenamefont {Mohamed},\ and\
  \citenamefont {Parrinello}}]{TDKuhne07}%
  \BibitemOpen
  \bibfield  {author} {\bibinfo {author} {\bibfnamefont {T.~D.}\ \bibnamefont
  {K\"{u}hne}}, \bibinfo {author} {\bibfnamefont {M.}~\bibnamefont {Krack}},
  \bibinfo {author} {\bibfnamefont {F.~R.}\ \bibnamefont {Mohamed}}, \ and\
  \bibinfo {author} {\bibfnamefont {M.}~\bibnamefont {Parrinello}},\
  }\href@noop {} {\bibfield  {journal} {\bibinfo  {journal} {Phys. Rev. Lett.}\
  }\textbf {\bibinfo {volume} {98}},\ \bibinfo {pages} {066401} (\bibinfo
  {year} {2007})}\BibitemShut {NoStop}%
\bibitem [{\citenamefont {Atsumi}\ and\ \citenamefont
  {Nakai}(2008)}]{TAtsumi08}%
  \BibitemOpen
  \bibfield  {author} {\bibinfo {author} {\bibfnamefont {T.}~\bibnamefont
  {Atsumi}}\ and\ \bibinfo {author} {\bibfnamefont {H.}~\bibnamefont {Nakai}},\
  }\href@noop {} {\bibfield  {journal} {\bibinfo  {journal} {Comput. Phys.
  Commun.}\ }\textbf {\bibinfo {volume} {167}},\ \bibinfo {pages} {103}
  (\bibinfo {year} {2008})}\BibitemShut {NoStop}%
\bibitem [{\citenamefont {Fang}\ \emph {et~al.}(2016)\citenamefont {Fang},
  \citenamefont {Gao}, \citenamefont {Song},\ and\ \citenamefont
  {Wang}}]{JFang16}%
  \BibitemOpen
  \bibfield  {author} {\bibinfo {author} {\bibfnamefont {J.}~\bibnamefont
  {Fang}}, \bibinfo {author} {\bibfnamefont {X.}~\bibnamefont {Gao}}, \bibinfo
  {author} {\bibfnamefont {H.}~\bibnamefont {Song}}, \ and\ \bibinfo {author}
  {\bibfnamefont {H.}~\bibnamefont {Wang}},\ }\href@noop {} {\bibfield
  {journal} {\bibinfo  {journal} {J. Chem. Phys}\ }\textbf {\bibinfo {volume}
  {144}},\ \bibinfo {pages} {244103} (\bibinfo {year} {2016})}\BibitemShut
  {NoStop}%
\bibitem [{\citenamefont {Coretti}\ \emph {et~al.}(2018)\citenamefont
  {Coretti}, \citenamefont {Bonella},\ and\ \citenamefont
  {Ciccotti}}]{ACoretti18}%
  \BibitemOpen
  \bibfield  {author} {\bibinfo {author} {\bibfnamefont {A.}~\bibnamefont
  {Coretti}}, \bibinfo {author} {\bibfnamefont {S.}~\bibnamefont {Bonella}}, \
  and\ \bibinfo {author} {\bibfnamefont {G.}~\bibnamefont {Ciccotti}},\
  }\href@noop {} {\bibfield  {journal} {\bibinfo  {journal} {J. Chem. Phys.}\
  }\textbf {\bibinfo {volume} {149}},\ \bibinfo {pages} {191102} (\bibinfo
  {year} {2018})}\BibitemShut {NoStop}%
\bibitem [{\citenamefont {Bonella}\ \emph {et~al.}(2020)\citenamefont
  {Bonella}, \citenamefont {Coretti}, \citenamefont {Vuilleumier},\ and\
  \citenamefont {Ciccotti}}]{SBonella20}%
  \BibitemOpen
  \bibfield  {author} {\bibinfo {author} {\bibfnamefont {S.}~\bibnamefont
  {Bonella}}, \bibinfo {author} {\bibfnamefont {A.}~\bibnamefont {Coretti}},
  \bibinfo {author} {\bibfnamefont {R.}~\bibnamefont {Vuilleumier}}, \ and\
  \bibinfo {author} {\bibfnamefont {G.}~\bibnamefont {Ciccotti}},\ }\href@noop
  {} {\bibfield  {journal} {\bibinfo  {journal}
  {https://arxiv.org/pdf/2001.03556.pdf}\ } (\bibinfo {year}
  {2020})}\BibitemShut {NoStop}%
\bibitem [{\citenamefont {Payne}\ \emph {et~al.}(1992)\citenamefont {Payne},
  \citenamefont {Teter}, \citenamefont {Allan}, \citenamefont {Arias},\ and\
  \citenamefont {Joannopoulos}}]{MCPayne92}%
  \BibitemOpen
  \bibfield  {author} {\bibinfo {author} {\bibfnamefont {M.~C.}\ \bibnamefont
  {Payne}}, \bibinfo {author} {\bibfnamefont {M.~P.}\ \bibnamefont {Teter}},
  \bibinfo {author} {\bibfnamefont {D.~C.}\ \bibnamefont {Allan}}, \bibinfo
  {author} {\bibfnamefont {T.~A.}\ \bibnamefont {Arias}}, \ and\ \bibinfo
  {author} {\bibfnamefont {J.~D.}\ \bibnamefont {Joannopoulos}},\ }\href@noop
  {} {\bibfield  {journal} {\bibinfo  {journal} {Rev. Mod. Phys.}\ }\textbf
  {\bibinfo {volume} {64}},\ \bibinfo {pages} {1045} (\bibinfo {year}
  {1992})}\BibitemShut {NoStop}%
\bibitem [{\citenamefont {Arias}\ \emph {et~al.}(1992)\citenamefont {Arias},
  \citenamefont {Payne},\ and\ \citenamefont {Joannopoulos}}]{TAArias_92}%
  \BibitemOpen
  \bibfield  {author} {\bibinfo {author} {\bibfnamefont {T.}~\bibnamefont
  {Arias}}, \bibinfo {author} {\bibfnamefont {M.}~\bibnamefont {Payne}}, \ and\
  \bibinfo {author} {\bibfnamefont {J.}~\bibnamefont {Joannopoulos}},\
  }\href@noop {} {\bibfield  {journal} {\bibinfo  {journal} {Phys. Rev. B}\
  }\textbf {\bibinfo {volume} {45}},\ \bibinfo {pages} {1538} (\bibinfo {year}
  {1992})}\BibitemShut {NoStop}%
\bibitem [{\citenamefont {Barnett}\ and\ \citenamefont
  {Landman}(1993)}]{RBarnett93}%
  \BibitemOpen
  \bibfield  {author} {\bibinfo {author} {\bibfnamefont {R.~N.}\ \bibnamefont
  {Barnett}}\ and\ \bibinfo {author} {\bibfnamefont {U.}~\bibnamefont
  {Landman}},\ }\href@noop {} {\bibfield  {journal} {\bibinfo  {journal} {Phys.
  Rev. B}\ }\textbf {\bibinfo {volume} {48}},\ \bibinfo {pages} {2081}
  (\bibinfo {year} {1993})}\BibitemShut {NoStop}%
\bibitem [{\citenamefont {Kresse}\ and\ \citenamefont
  {Hafner}(1993)}]{GKresse93}%
  \BibitemOpen
  \bibfield  {author} {\bibinfo {author} {\bibfnamefont {G.}~\bibnamefont
  {Kresse}}\ and\ \bibinfo {author} {\bibfnamefont {J.}~\bibnamefont
  {Hafner}},\ }\href@noop {} {\bibfield  {journal} {\bibinfo  {journal} {Phys.
  Rev. B}\ }\textbf {\bibinfo {volume} {47}},\ \bibinfo {pages} {558} (\bibinfo
  {year} {1993})}\BibitemShut {NoStop}%
\end{thebibliography}%
\end{document}